\documentclass[ onecolumn]{IEEEtran}

   \usepackage{nopageno}
\usepackage{amsmath,stmaryrd,acronym,amssymb,amsthm,dsfont,graphicx}
\graphicspath{{graphics/}}

\acrodef{AEP}{Asymptotic Equipartition Property}
\acrodef{AoA}{Angle of Arrival}
\acrodef{AWGN}{Additive White Gaussian Noise}
\acrodef{BER}{Bit-Error-Rate}
\acrodef{BEC}{Binary Erasure Channel}
\acrodef{BPSK}{Binary Phase-Shift Keying}
\acrodef{BSC}{Binary Symmetric Channel}
\acrodef{CDF}[CDF]{Cumulative Distribution Function}
\acrodef{CLT}[CLT]{Central Limit Theorem}
\acrodef{CSI}[CSI]{Channel State Information}
\acrodef{DMC}[DMC]{Discrete Memoryless Channel}
\acrodef{DMS}[DMS]{Discrete Memoryless Source}
\acrodef{iid}[i.i.d.]{independent and identically distributed}
\acrodef{lhs}[l.h.s.]{left-hand-side}
\acrodef{rhs}[r.h.s.]{right-hand-side}
\acrodef{LPD}[LPD]{Low Probability of Detection}
\acrodef{LDPC}[LDPC]{Low-Density Parity-Check}
\acrodef{MAC}[MAC]{multiple-access channel}
\acrodef{MIMO}[MIMO]{Multiple-Input Multiple-Output}
\acrodef{MISO}{Multiple-Input Single-Output}
\acrodef{PDF}[PDF]{Probability Distribution Function}
\acrodef{PMF}[PMF]{Probability Mass Function}
\acrodef{PPM}[PPM]{Pulse Position Modulation}
\acrodef{PSD}{Power Spectral Density}
\acrodef{QPSK}{Quadrature Phase-Shift Keying}
\acrodef{SIMO}{Single-Input Multiple-Output}
\acrodef{SNR}{Signal-to-Noise Ratio}
\acrodef{wrt}[w.r.t.]{with respect to}
\acrodef{WSS}{Wide Sense Stationary}

\DeclareMathAlphabet{\eurm}{U}{eur}{m}{n}
\DeclareMathAlphabet{\mathbsf}{OT1}{cmss}{bx}{n}
\DeclareMathAlphabet{\mathssf}{OT1}{cmss}{m}{sl}
\DeclareMathAlphabet{\mathcsf}{OT1}{cmss}{sbc}{n}

\DeclareSymbolFont{bsfletters}{OT1}{cmss}{bx}{n}  
\DeclareSymbolFont{ssfletters}{OT1}{cmss}{m}{n}
\DeclareMathSymbol{\bsfGamma}{0}{bsfletters}{'000}
\DeclareMathSymbol{\ssfGamma}{0}{ssfletters}{'000}
\DeclareMathSymbol{\bsfDelta}{0}{bsfletters}{'001}
\DeclareMathSymbol{\ssfDelta}{0}{ssfletters}{'001}
\DeclareMathSymbol{\bsfTheta}{0}{bsfletters}{'002}
\DeclareMathSymbol{\ssfTheta}{0}{ssfletters}{'002}
\DeclareMathSymbol{\bsfLambda}{0}{bsfletters}{'003}
\DeclareMathSymbol{\ssfLambda}{0}{ssfletters}{'003}
\DeclareMathSymbol{\bsfXi}{0}{bsfletters}{'004}
\DeclareMathSymbol{\ssfXi}{0}{ssfletters}{'004}
\DeclareMathSymbol{\bsfPi}{0}{bsfletters}{'005}
\DeclareMathSymbol{\ssfPi}{0}{ssfletters}{'005}
\DeclareMathSymbol{\bsfSigma}{0}{bsfletters}{'006}
\DeclareMathSymbol{\ssfSigma}{0}{ssfletters}{'006}
\DeclareMathSymbol{\bsfUpsilon}{0}{bsfletters}{'007}
\DeclareMathSymbol{\ssfUpsilon}{0}{ssfletters}{'007}
\DeclareMathSymbol{\bsfPhi}{0}{bsfletters}{'010}
\DeclareMathSymbol{\ssfPhi}{0}{ssfletters}{'010}
\DeclareMathSymbol{\bsfPsi}{0}{bsfletters}{'011}
\DeclareMathSymbol{\ssfPsi}{0}{ssfletters}{'011}
\DeclareMathSymbol{\bsfOmega}{0}{bsfletters}{'012}
\DeclareMathSymbol{\ssfOmega}{0}{ssfletters}{'012}

\newcommand{\calA}{{\mathcal{A}}}
\newcommand{\calB}{{\mathcal{B}}}

\newcommand{\calD}{{\mathcal{D}}}
\newcommand{\calE}{{\mathcal{E}}}

\newcommand{\calH}{{\mathcal{H}}}

\newcommand{\calJ}{{\mathcal{J}}}
\newcommand{\calK}{{\mathcal{K}}}

\newcommand{\calM}{{\mathcal{M}}}
\newcommand{\calN}{{\mathcal{N}}}

\newcommand{\calP}{{\mathcal{P}}}

\newcommand{\calT}{{\mathcal{T}}}
\newcommand{\calS}{{\mathcal{S}}}

\newcommand{\calX}{{\mathcal{X}}}

\newcommand{\E}[2][]{{\mathbb{E}_{#1}}{\left(#2\right)}}       
\renewcommand{\P}[2][]{{\mathbb{P}_{#1}}{\left(#2\right)}}
       
\newcommand{\avgI}[1]{{{\mathbb{I}}\!\left(#1\right)}}
\newcommand{\avgH}[1]{{\mathbb{H}}\!\left(#1\right)}

\newcommand{\norm}[2][]{\ensuremath{{\left\Vert{#2}\right\Vert}_{#1}}}   
\newcommand{\eqdef}{\ensuremath{\triangleq}}                    
\newcommand{\intseq}[2]{\ensuremath{\llbracket{#1},{#2}\rrbracket}}  
\newcommand{\indic}[1]{\ensuremath{\mathds{1}\!\left\{#1\right\}}}

\renewcommand{\leq}{\leqslant}
\renewcommand{\geq}{\geqslant}

\newcommand{\tr}[1]{\ensuremath{\text{\textnormal{tr}}\left(#1\right)}}  

\newcommand{\proddist}{%
  \mathchoice{\raisebox{1pt}{$\displaystyle\otimes$}}
             {\raisebox{1pt}{$\otimes$}}
             {\raisebox{0.5pt}{\scalebox{0.7}{$\scriptstyle\otimes$}}}
             {\raisebox{0.4pt}{\scalebox{0.6}{$\scriptscriptstyle\otimes$}}}}
\newcommand{\pn}{{\proddist n}}

\usepackage[textwidth=0.68in,textsize=footnotesize,disable]{todonotes}
\presetkeys{todonotes}{color=redArcom, linecolor=redArcom,bordercolor=redArcom}{}

\newtheorem{theorem}{Theorem}
\newtheorem{remark}{Remark}
\newtheorem{definition}{Definition}
\newtheorem{lemma}{Lemma}
\newtheorem{corollary}{Corollary}
\newtheorem{proposition}{Proposition}

\acrodef{ROC}[ROC]{Receiver Operation Characteristic}
\acrodef{PPM}[PPM]{Pulse-Position Modulation}

\newcommand{\pr}[1]{{\left(#1\right)}}
\newcommand{\set}[1]{{\left\{#1\right\}}}

\presetkeys{todonotes}{color=redArcom, linecolor=redArcom,bordercolor=redArcom}{}

\acrodef{AVC}{arbitrary varying channel}

\usepackage{diagbox}

\newcommand{\ket}[1]{|#1\rangle}
\newcommand{\bra}[1]{\langle #1 |}
\newcommand{\braket}[2]{\langle #1 | #2 \rangle}
\newcommand{\id}{\mathrm{id}}
\newcommand{\cid}{\overline{\mathrm{id}}}
\newcommand{\one}{\mathbf{1}}

\newcommand{\N}{\mathbb{N}}

\newcommand{\tA}{{\widetilde{A}}}
\newcommand{\tB}{{\widetilde{B}}}
\newcommand{\wc}{{w}}
\newcommand{\mc}{{M}}
\newcommand{\ws}{{\overline{w}}}
\newcommand{\ms}{{\overline{M}}}
\newcommand{\ks}{{\overline{K}}}
\newcommand{\ih}[3]{{\aleph^{#1}\pr{#2}_{#3}}}
\newcommand{\ir}[3]{{\beth^{#1}\pr{#2}_{#3}}}
\usepackage{tikz}

\newcommand{\kb}[1]{{\ket{#1}\bra{#1}}}
\newcommand{\ptr}[2]{\mathrm{tr}_{#1}\pr{#2}}
\begin{document}

\title{Steganography Protocols for Quantum Channels}
\author{Mehrdad Tahmasbi and Matthieu Bloch \thanks{ This work was supported by NSF under award TWC 1527387.}}
\maketitle
\begin{abstract}
We study several versions of a quantum steganography problem, in which two legitimate parties attempt to conceal a cypher in a quantum cover transmitted over a quantum channel without arising suspicion from a warden who intercepts the cover. In all our models, we assume that the warden has an inaccurate knowledge of the quantum channel and we formulate several variations of the steganography problem depending on the tasks used as the cover and the cypher task. In particular, when the cover task is classical communication, we show that the cypher task can be classical communication or entanglement sharing; when the cover task is entanglement sharing and the main channel is noiseless, we show that the cypher task can be randomness sharing; when the cover task is quantum communication and the main channel is noiseless, we show that the cypher task can be classical communication. In the latter case, our results improve earlier ones by relaxing the need for a shared key between the transmitter and the receiver and hold under milder assumptions on the cover quantum communication code. 
\end{abstract}

\section{Introduction}
\label{sec:introduction}
In steganography, two parties seek to  embed  information  within an innocent looking message without being detected by an unwanted party. The well-known example is that of two prisoners, Alice and Bob, who aim at developing an escape plan (\emph{cyphertext}) through a permissible communication (\emph{covertext}). The resulting message (\emph{stegotext}), which is a combination of cyphertext, covertext, and possibly of a shared secret key, shall be made available to a warden Willie and should be almost indistinguishable from the covertext. While this fictional example illustrates the main motivation behind the problem,  the advent of the digital age has opened  several real opportunities to conceal information, including the embedding of messages in digital images and texts as well as telecommunication networks. Applications of modern steganography are now numerous and range from copyright protection  to malicious activities. The importance of such applications has led to the formalization of steganography using sound cryptographic principles and the development of both steganography methods and their countermeasures \cite{Cox2007}.

The classical information-theoretic limits of information-hiding and steganography have been studied using different measures of ``hiding.'' The measures include average distortion between the covertext and the stegotext~\cite{Moulin2003,Wang2008} as well as relative entropy between the distributions of the covertext and stegotext~\cite{Cachin2004,Ker2007}, which essentially controls the performance of the warden's optimal detector. More recently, these ideas have also been applied in the context of covert and stealth communications~\cite{Bash2013,Hou2014}. The main insight derived from these works is the precise characterization of the number of covert bits that can be embedded in the covertext while remaining undetectable by Willie and of the number of secret key bits required by Alice and Bob to achieve this goal. The number of covert bits is sensitive to modeling assumptions, in particular to whether Willie knows the covertext or whether there is noise in the system. The authors of \cite{Sobers2017} have shown that  reliable and covert transmission of $O(n)$ bits of information is possible in $n$ uses of an \ac{AWGN} channel when the warden has uncertainty about  the noise power of the channel. The authors of \cite{Soltani2018, Shahrzad2018} have moreover considered  covert communication when friendly nodes transmit artificial noise and have proved that  covert transmission of positive rates is possible. Another situation in which covert communication with positive rate was shown to be possible is  the transmission from a relay node to a destination when  the source is uncertain regarding the forwarding strategy of the relay node \cite{Hu2018}.

Concurrently, the quantum description of  physical devices used in information processing tasks has made us   re-think   communication and computation problems from two perspectives. First, one can  use the limits imposed by quantum mechanics to devise enhanced solutions to  hard problems in the classical world. For example, quantum key distribution offers unconditional security for classical communication while most classical solutions rely on assumptions regarding the computational power of the adversary. Second,  one often encounters new challenging problems in a quantum setting, such as entanglement generation, which  plays a role in intriguing  applications such as quantum teleportation and super dense coding. Returning to the problem of steganography, one can extend the classical formulation to encompass both these aspects. That is, in addition to leveraging the quantum nature of the communication channel to perform classical steganography, one can ask for new paradigms to hide various quantum information processing tasks. Alice and Bob could for instance conceal a classical message within a quantum error correcting code used to mitigate the quantum noise of a quantum computer. Because of the unique nature of quantum states and channels, quantum steganography is in principle richer than classical steganography~\cite{Natori2006}, and much efforts have been devoted to characterize how much information can be embedded into various quantum channels with or without noise \cite{Gea-Banacloche2002,Shaw2011,Sheikholeslami2016,Sutherland2018, Sanguinetti2016,Sutherland2018a}, and to assess how much key is required to achieve the task.

We revisit here the model of quantum steganography put forward in~\cite{Sutherland2018a,Sutherland2018}, which \emph{assumes that the warden has inaccurate knowledge of what the channel is}. Specifically, we assume that the warden's knowledge of the channel is a degraded version of the real channel, which can be achieved by intentionally cascading another channel at the transmitter.  We develop and analyze several quantum steganography protocols and obtain the following four  results summarized in Table~\ref{tb:results}.\footnote{Please note that item 1, 3, and 4  are included in the conference version \cite{Tahmasbi2019} without detailed proofs.}
\begin{table}[]
\caption{ {\color{green}\checkmark} noisy main channel\quad{\color{blue}\checkmark} noiseless main channel}

\centering
\begin{tabular}{|c|ccc|}
\hline
\diagbox[width=6em]{Cypher}{$~$ Cover} &CC&ES&QC\\ 
\hline
 CC  & {\color{green}\checkmark} &    & {\color{blue}\checkmark}   \\
 CRS&  &  {\color{blue}\checkmark}  &  \\
 ES&  {\color{green}\checkmark}&    &  \\ 
  \hline
\end{tabular}

\label{tb:results}


\end{table}
\begin{enumerate}
\item When the cover protocol consists in communicating classically over a quantum channel, we show that, in addition to the cover classical message, a cypher classical message can be transmitted (Theorem~\ref{th:main-cc}).
\item When the cover protocol consists in communicating classically over a quantum channel, we show that, in addition to the cover classical message, entangled qubits can be generated. (Theorem~\ref{th:main-cc-es}).
\item When the cover protocol consists in sharing entanglement and the channel is noiseless, we show that  legitimate parties can  share entanglement as well as classical randomness (Theorem~\ref{th:main-es}).
\item When the cover protocol consist of a quantum communication and the channel is noiseless, we show that, in addition to the cover quantum message, a cypher classical message can be transmitted (Theorem~\ref{th:main-qc}).
\end{enumerate}
In all aforementioned results, the observed channel output state  when the stego protocol is executed over the true channel   resembles the observed state when the cover protocol is executed over the channel expected by the warden. Unlike earlier results  \cite{Shaw2011,Sutherland2018a,Sutherland2018}, we show that  no shared key is required to run the stego protocol when the channel is noiseless. This is achieved through the use of a random encoder obtained from privacy amplification and source coding with side information techniques similar to \cite{Renes2011, Yassaee2014}. Furthermore, we relax the assumption on the cover code in \cite{Sutherland2018a} that ``\emph{on a valid codeword in the QECC, the typical errors  all have distinct error syndromes, and act as unitaries that move the state to a distinct, orthogonal subspace},'' by relying on one-shot coding results. Our main results are not single-letterized because of the arbitrary structure of the cover code; however, we specialize our results to certain classes of codes and obtain single-letter expression for those examples.

The remainder of the paper is organized as follows. We introduce our notation in Section~\ref{sec:notation}. We formulate different information process protocols over a quantum channel and define our problem in Section~\ref{sec:prob-form}. We state our main theorems in Section~\ref{sec:main-res}. We next calculate the rate of the cypher protocol for specific instances of cover protocols in Section~\ref{sec:example}. We finally prove the main theorems in Section~\ref{sec:proof}.
\section{Notation}
\label{sec:notation}
We assume that all systems (e.g., $A$) are described by finite-dimensional Hilbert spaces (e.g., $\calH_A$). Let $\one_A$ be the identity map on $\calH_A$. $\calB(\calH_A)$ denotes the set of all bounded linear operators from $\calH_A$ to $\calH_A$, $\calP(\calH_A)$ denotes the set of all positive operators in $\calB(\calH_A)$, and $\calD(\calH_A)$ denotes the set of  all density operators on $\calH_A$. For $X\in \calB(\calH_A)$, the trace norm of $X$ is ${\norm{X}}_1 \eqdef \tr{\sqrt{X^\dagger X}}$, and $\nu(X)$ denotes the number of \emph{distinct} eigenvalues of $X$. The fidelity between two density operators $\rho_A$ and $\sigma_A$ is defined as $F(\rho_A, \sigma_A) \eqdef {\norm{\sqrt{\rho_A} \sqrt{\sigma_A}}}_1^2$.  A quantum channel $\calN_{A\to B}$ is a linear trace-preserving completely positive map from $\calB(\calH_A)$ to $\calB(\calH_B)$. Let $\id_A$ be the identity channel on $\calB(\calH_A)$ and $\emptyset_A$ be the channel that maps all states in $\calD(\calH_A)$ to the trivial state in a one-dimensional state. 

Suppose that $\rho_{XB} = \sum_{x}P_X(x) \ket{x}\bra{x} \otimes \rho_B^x$ is a classical-quantum (cq) state. We recall two versions of R\'enyi quantum mutual information \cite{Hayashi2006} for $a\neq 0$,
\begin{align}
\ih{a}{X;B}{\rho} &\eqdef  -\frac{1}{a} \log\pr{\tr{\rho_{XB}(\rho_X\otimes \rho_B)^{\frac{a}{2}} \rho_{XB}^{-a}(\rho_X\otimes \rho_B)^{\frac{a}{2}} } },\displaybreak[0]\\
\ir{a}{X;B}{\rho} &\eqdef - \frac{1}{a} \log\pr{\sum_x P_X(x) \tr{(\rho_B^x)^{1-a}\rho_B^a}} .
\end{align}
We also define the R\'enyi quantum entropy as  $H^a(\rho) \eqdef -\frac{1}{a} \log \tr{\rho^{a+1}}$ \cite{Hayashi2006}.
These quantities are approximated by the Holevo information when $\rho$ and $\calN$ have a product structure and are useful to express the coding theorems for cq channels~\cite{Hayashi2006,Tomamichel2013, Wang2012}.

For a positive integer $M$, let $\calH^{(M)}$ denote the $M$-dimensional space spanned by the orthonormal basis $\set{\ket{1}, \cdots, \ket{M}}$. We also define  $\id^{(M)}(\rho) \eqdef \rho$ and $\cid^{(M)}(\rho) \eqdef \sum_{i=1}^M \kb{i} \rho \kb{i}$ for $\rho\in \calD\pr{\calH^{(M)}}$. Furthermore, we define the perfectly entangled and the perfectly classically correlated states
\begin{align}
\ket{\Phi^{(M)}} &\eqdef \frac{1}{\sqrt{M}} \sum_{i=1}^M \ket{i} \otimes \ket{i} \in \calH^{(M)} \otimes \calH^{(M)}\\
\overline{\Phi}^{(M)} &\eqdef \frac{1}{M} \sum_{i=1}^M \kb{i} \otimes \kb{i} \in \calD\pr{\calH^{(M)} \otimes \calH^{(M)}}.
\end{align}
\section{Problem Formulation}
\label{sec:prob-form}

Suppose that Alice and Bob are connected by a quantum channel $\calN_{A\to B}:\calB(\calH_A)\to \calB(\calH_B)$ and use the channel $n$ times to run a protocol, which could be a combination of four primary tasks  (classical communication, quantum communication, randomness sharing, and entanglement sharing), as defined next.

\begin{itemize}
\item \textbf{Classical Communication:} Alice wishes to reliably transmit a classical message $W$ uniformly distributed over $\intseq{1}{M}$. A code consists of a function $f: \intseq{1}{M}\to \calD(\calH_A^\pn)$ for Alice to encode message $w$ into an input state $\rho_{A^n}^w\eqdef f(w)$ and a POVM $\mathbf{\Lambda} = \{\Lambda^w\}_{w\in\intseq{1}{M}}$ for Bob to decode $W$. We call the code an $(M, \epsilon)^{\textnormal{CC}}$ classical communication code, if we have $\frac{1}{M}\sum_{w=1}^M \tr{\Lambda^w \calN^{\pn}_{A\to B}(f(w))} \geq 1-\epsilon$.
The induced output state is $ \frac{1}{M}\sum_{w=1}^M \calN_{A\to B}^\pn(f(w))$.
\item \textbf{Quantum Communication:} Alice wants to transmit a quantum state $\rho_W$ acting on an $M$-dimensional Hilbert space $\calH_W \eqdef \calH^{(M)}$. Alice encodes $\rho_W$ using an encoder $\calE_{W\to A^n}$ and transmits it over $n$ uses of $\calN_{A\to B}$. Bob decodes $\rho_W$ by applying a decoder $\calD_{B^n\to W}$ to his received state. A code $(\calE_{W\to A^n}, \calD_{B^n\to W})$ is an  $(M, \epsilon)^{\textnormal{QC}}$ code if
\begin{align}
\min_{\rho_W \in \calD(\calH_W)} F(\rho_W, (\calD_{B^n\to W}\circ \calN_{A\to B}^\pn \circ\calE_{W\to A^n} )(\rho_W)) \geq 1-\epsilon.
\end{align}
A more stringent notion of reliability is that the code recovers most of the error operators applied by the channel. Formally, we call a code $(\calE_{W\to A^n}, \calD_{B^n \to B})$ an $(M, \epsilon)^{\textnormal{QC}}_{\textnormal{R}}$ code, if there exists a decomposition $\calN_{A\to B}^\pn = \widetilde{\calN}_{A^n\to B^n} + \widetilde{\widetilde{\calN}}_{A^n\to B^n}$ such that $\calD_{B^n \to W} \circ \widetilde{\calN}_{A^n \to B^n} \circ \calE_{W\to A^n} = c~\id_W$ for $c\geq 1-\epsilon$.
The induced output state is $\calN_{A\to B}^\pn(\calE_{W\to A^n}(\rho_W))$ when the message is $\rho_W$.

\item \textbf{Randomness Sharing:} Alice and Bob desire to share a classical random variable $\overline{\Phi}^{(M)}$. Let $\calH_{\tA} = \calH_{\tB} \eqdef \calH^{(M)}$. Alice prepares a state $\rho_{\widetilde{A} A^n}$ over the Hilbert space $\calH_{\widetilde{A}}\otimes \calH_A^\pn$ and transmits $\rho_{A^n}$ to Bob over $n$ uses of the channel $\calN_{A\to B}$. Bob applies a decoder $\calD_{B^n\to \widetilde{B}}$ to his received state $\calN_{A\to B}^\pn(\rho_{A^n})$ to obtain the state $\rho_{\widetilde{B}}$ acting on the Hilbert space $\calH_{\widetilde{B}}$. The joint state $\rho_{\tA\tB} \eqdef (\id_\tA\otimes (\calD_{B^n\to \tB} \circ \calN_{A\to B} ^\pn)) (\rho_{\tA A^n})$ is their final shared randomness. A code $(\rho_{\tA A^n}, \calD_{B^n\to \tB})$ is called an $(M, \epsilon)^{\textnormal{RS}}$  randomness sharing code if $ F(\overline{\Phi}^{(M)}, \rho_{\tA \tB}) \geq 1-\epsilon$. The induced output state is $\calN_{A\to B}^\pn(\rho_{A^n})$.

\item \textbf{Entanglement Sharing:} Alice and Bob want to share the entangled state $\Phi^{(M)}$. An $(M, \epsilon)^{\textnormal{ES}}$ code is defined in the same way as a  randomness sharing protocol except that the final desired state is $\Phi^{(M)}$. The induced output state is defined similarly to that of randomness sharing.
\end{itemize}

In the resource framework formulated in \cite{devetak2008resource}, these four protocols correspond to the simulation of $\cid^{(M)}$, $\id^{(M)}$, $\overline{\Phi}^{(M)}$, and $\Phi^{(M)}$ with $n$ uses of $\calN_{A\to B}$. Alice and Bob can in principle desire to perform any combination of these four protocols over $n$ uses of the channel $\calN_{A\to B}$. We formalize only the combinations for which we develop  results, i.e., classical communication / quantum communication, entanglement sharing / randomness sharing, and entanglement sharing / classical communication.

\begin{itemize}

\item \textbf{Quantum and Classical Communication:} Alice wants to transmit a quantum state $\rho_W$ over an $M$-dimensional space $\calH_W \eqdef \calH^{(M)}$ and an independent classical message $\overline{W}$ uniformly distributed over $\intseq{1}{\ms}$. When $\overline{W} = \ws$, she encodes $\rho_W$ using the encoder $\calE^{\ws}_{W\to A^n}$. Bob decodes the messages using a decoder $\calD_{B^n \to W\overline{W}}$. The code is called an $(M, \overline{M}, \epsilon)^{\textnormal{QC-CC}}$ code if for any $\rho_W$, we have
\begin{align}
\frac{1}{\overline{M}} \sum_{\ws} \tr{\ket{\ws}\bra{\ws} (\calD_{B^n\to \overline{W}} \circ\calN_{A\to B}^\pn \circ \calE^\ws_{W\to A^n})(\rho_W)} \geq 1-\epsilon,
\end{align}
and for all $\ws\in \intseq{1}{\ms}$, $(\calE_{W\to A^n}^\ws, \calD_{B^n \to W})$ is an $(M, \epsilon)^{\textnormal{QC}}_R$ code. The induced output state is $\frac{1}{\ms}\sum_{\ws = 1}^\ms\calN_{A\to B}^\pn(\calE^\ws_{W\to A^n}(\rho_W))$ when the quantum message is $\rho_W$.

\item \textbf{Entanglement and Randomness Sharing:} Alice and Bob want to share the state $\Phi^{(M)}\otimes \smash{\overline{\Phi}}^{(\overline{M})}$. An $(M, \overline{M}, \epsilon)^{\textnormal{ES-RS}}$ code is defined in the same way as a  randomness sharing protocol except that the final desired state is $\Phi^{(M)}\otimes \smash{\overline{\Phi}}^{(\overline{M})}$. The induced output state is defined similarly to that of randomness sharing.
\item \textbf{Classical Communication and Entanglement Sharing}: Alice wants to transmit a classical message $W$ uniformly distributed over $\intseq{1}{\mc}$ and share the entangled state $\Phi^{(\ms)}$ with Bob. Let $\calH_{\tA} = \calH_{\tB} \eqdef \calH^{(\ms)}$ and $\calH_W \eqdef \calH^{(\mc)}$. A code consists of an encoder $f:\intseq{1}{\mc} \to \calH_{\tA} \otimes \calH_{A^n}$ and a decoder $\calD_{B^n\to W\tB}$. Given the classical message $W = w$, Alice prepares $f(w)$ and sends the subsystem $A^n$ over $n$ uses of $\calN_{A\to B}$. Bob applies $\calD_{B^n \to W \tB}$ to his received state. We call $(f, \calD_{B^n\to W\tB})$ an $(\mc, \ms, \epsilon)^{\mathrm{CC-ES}}$ code if
\begin{align}
\frac{1}{\mc} \sum_{\wc}  \bra{\wc} \calD_{B^n \to W} \circ \calN_{A\to B}^\pn (\ptr{\tA}{f(\wc)}) \ket{\wc} \geq 1-\epsilon,\\
{\norm{\frac{1}{\mc}\sum_{\wc} (\id_{\tA} \otimes (\calD_{B^n\to \tB}\circ \calN_{A\to B}^\pn)) (f(w))  -\Phi^{(\ms)}}}_1 \leq \epsilon.
\end{align}
The induced output state is $\frac{1}{M}\sum_{w=1}^M \calN_{A\to B}^\pn( \mathrm{tr}_{\tA}(f(w)))$.

\end{itemize}
All these protocols can be enhanced with a shared secret classical key $S$ uniformly distributed over $\intseq{1}{K}$, which can  help Alice and Bob  induce a specific output state.
\begin{figure}[t]
  \centering
  \includegraphics[width=.4\linewidth]{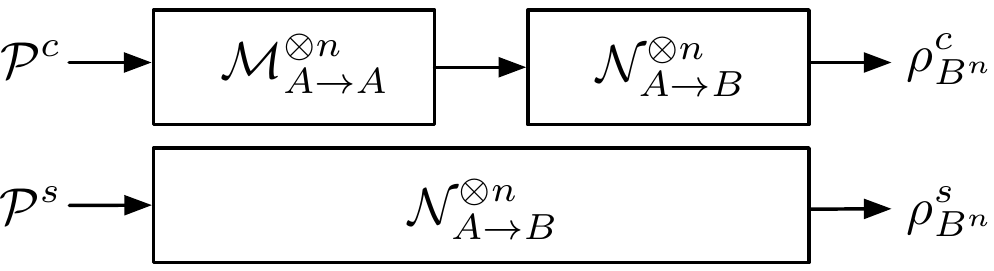}
  \caption{Willie's expectation (top) and true communication (bottom)}
    \label{fig:model}

\end{figure} 
As depicted in Fig.~\ref{fig:model}, Willie expects Alice and Bob to execute a protocol $\calP^c$, which is called \emph{the cover protocol} and is known to Willie. However, Willie has an  inaccurate estimation of the channel and believes that the channel between Alice and Bob is $\calN_{A\to B} \circ \calM_{A\to A}$, which is a degraded version of the true channel $\calN_{A\to B}$. We assume that running the protocol $\calP^c$ induces the quantum state $\rho_{B^n}^c$ at the output of $\calN_{A\to B}^\pn\circ\calM_{A\to A}^\pn$. The objective is for Alice and Bob  to run a \emph{stego protocol} $\calP^s$, which performs the task of $\calP^c$ together with another task and induces a state $\rho_{B^n}^s$ at the output of $\calN_{A\to B}^\pn$ such that ${\norm{\rho_{B^n}^c - \rho_{B^n}^s}}_1$ is small. The added tasks can be any of the tasks listed earlier. 
We focus on four of these as summarized in Table~\ref{tb:results} and detailed next.

\section{Main Results}
\label{sec:main-res}
We state our main results in this section, and all proofs are relegated to Section~\ref{sec:proof}.
We first show that if the cover protocol is a classical communication code, the stego protocol could be a classical communication code with a higher rate,  equivalent to sending a cypher classical message in addition to the cover classical message.

\begin{theorem}[classical communication / classical communication]
  
\label{th:main-cc}
 Let the cover protocol be an $(\mc, \epsilon)^{\textnormal{CC}}$ code $(f, \mathbf{\Lambda})$ for  $\calN_{A\to B}^\pn \circ \calM_{A\to A}^\pn$ inducing the output state $\rho_{B^n}^c$. We define $\rho_{B^n}^\wc \eqdef \calN_{A\to B}^\pn \circ \calM_{A\to A}^\pn(f(\wc))$ for $\wc\in\intseq{1}{\mc}$.
\begin{itemize}
\item Suppose that $\calH_A = \calH_B$ and $\calN_{A\to B} = \id_A$, i.e., the true channel from Alice to Bob is noiseless. For any $\zeta > 0$,  there exists an $(\mc\ms, \zeta+2\sqrt{\zeta+\epsilon})^{\textnormal{CC}}$ stego protocol inducing the output state $\rho_{B^n}^s$  such that $ {\norm{\rho_{B^n}^s - \rho_{B^n}^c}}_1 \leq \zeta$ provided that 
\begin{align}
\label{eq:cc-rate-noiseless}
\log \ms \leq  \min_{\wc\in\intseq{1}{\mc}}\sup_{a\in]0, 1[} H^a(\rho_{B^n}^\wc)  - \frac{4}{a} \log \frac{2}{\zeta}.
\end{align}

\item Suppose that the channel $\calN_{A\to B}$ is noisy. Let $\sigma_{XA^n}^1, \cdots, \sigma_{XA^n}^\mc$ be cq states such that upon defining $\sigma_{XB^n}^w \eqdef (\id_X \otimes \calN_{A\to B}^\pn)(\sigma_{XA^n}^\wc)$, we have $\textnormal{tr}_X(\sigma_{XB}^w)=\rho_{B^n}^\wc$ for all $\wc\in\intseq{1}{\mc}$.  Let $\zeta, \xi\in ]0, 1[$ be fixed and $\ms$ and $\ks$ be positive integers such that $\log \ms \leq \min_{\wc} \log \ms_\wc$ where
\begin{align}
\label{eq:cc-rate-noisy}
\log \ms_\wc \eqdef \sup_{a\in]0, 1[} \left[\ih{a}{X;B^n}{\sigma^\wc}-\frac{1}{a}\log \nu(\sigma_X^\wc\otimes \sigma_{B^n}^\wc) - \frac{4}{a}\log \frac{12}{\zeta}\right] 
\end{align}
and
\begin{align}
\label{eq:cc-key-rate-noisy}
\log \ks \geq \max_\wc\left\{ \inf_{a<0} \left[\ir{a}{X;B^n}{\sigma^\wc} + \log \nu(\rho_{B^n}^w) +\pr{2-  \frac{2}{a}} \log \frac{12}{\xi} \right] - \log \ms_w + 1\right\}.
\end{align}
There exists an $(\mc\ms,  \zeta + 2\sqrt{\xi + \epsilon})^{\textnormal{CC}}$ code with $\log \ks$ bits of required common randomness inducing the output state $\rho_{B^n}^s$ such that ${\norm{\rho_{B^n}^s - \rho_{B^n}^c}}_1 \leq\xi$.
\end{itemize} 
\end{theorem}
\begin{remark}
We assume for simplicity that the cover message is uniformly distributed, but the proof holds for all distributions on the cover message.
\end{remark}
\begin{remark}
The arbitrary choice of   $\sigma_{XA^n}^1, \cdots, \sigma_{XA^n}^\mc$ in the second part of Theorem~\ref{th:main-cc} is an essential part of most of the channel coding results, for example the choice of the channel input state  in the definition of the Holevo information of a quantum channel \cite[Definition 13.3.1]{wilde2013quantum}. We need however an additional requirement $\textnormal{tr}_X(\sigma_{XB}^w)=\rho_{B^n}^\wc$ to control the channel output statistics. 
\end{remark}
We next show that if the cover protocol is a classical communication code, we can use a stego protocol to share entanglement and communicate classically. We introduce the following two definitions to express our results. In the first definition we introduce a shorthand for the result of Theorem~\ref{th:main-cc}. It shall help us compactly state the next theorem as we use the stego protocol of Theorem~\ref{th:main-cc} as a sub protocol in our stego protocol of Theorem~\ref{th:main-cc-es}.
\begin{definition}
Let us fix $ \xi = \zeta$ in the second part of Theorem~\ref{th:main-cc}. For an encoder $f$ and positive number $\zeta$, let $\log \ms^{\mathrm{CC}}(f, \zeta)$ and $\log K^{\mathrm{CC}}(f, \zeta)$ be the  number of bits of the cypher message and the number of required key bits, respectively, in the stego protocol of Theorem~\ref{th:main-cc}. Note that these quantities are well-defined, because the right hand side of \eqref{eq:cc-rate-noiseless}, \eqref{eq:cc-rate-noisy}, and \eqref{eq:cc-key-rate-noisy} only depends on $f$, $\zeta$, and $\xi$ when the channel is fixed.
\end{definition}
We next introduce a notation for the maximum amount of entanglement that can be distilled from an arbitrary shared quantum state using local operations and classical communication, known as the entanglement distillation problem.
\begin{definition}
Let Alice and Bob share $\rho_{AB}$ and $\calH_{\tA} = \calH_{\tB} \eqdef \calH^{(M)}$. An entanglement distillation protocol consists of an encoder $\calE_{A\to C \tA}$ and a decoder $\calD_{BC \to \tB}$ such that the output of $\calE_{A \to C \tA}$ is always a cq state. Alice applies $\calE_{A\to C \tA}$ to $\rho_A$ to obtain a cq state $\rho_{C\tA}$ and transmits $C$  to Bob over a noiseless channel. Bob applies $\calD_{CB\to \tB}$ to his subsystem $B$ and the received classical message $C$. The code $(\calE_{A\to C\tA}, \calD_{CB\to \tB})$ is called an $(M, L, \rho_{AB}, \epsilon)^{\mathrm{ED}}$ code if $\dim{\calH_C} = L$ and 
\begin{align}
{\norm{(\id_{\tA} \otimes \calD_{CB\to \tB})\circ(\calE_{A\to C\tA}\otimes \id_B )(\rho_{AB})  - \Phi^{(M)}}}_1 \leq \epsilon.
\end{align}
We further define  $E_d(\rho_{AB},  L, \epsilon) \eqdef \max \set{M: \exists~ (M, L, \rho_{AB}, \epsilon)^{\mathrm{ED}} \mathrm{ code} }$.
\end{definition}
 When $\rho_{AB}$ is pure, it is known \cite{Bennett1996} that $\lim_{\epsilon \to 0}\lim_{n\to \infty} \frac{\log E_d(\rho_{AB}^\pn,2^{\Theta(\log n)}, \epsilon)}{n} = \avgH{A}_\rho$.

\begin{theorem}[classical communication / entanglement sharing]
\label{th:main-cc-es}
 Let the cover protocol be an $(\mc, \epsilon)^{\textnormal{CC}}$ code $(f, \mathbf{\Lambda})$ for $\calN_{A\to B}^\pn \circ \calM_{A\to A}^\pn$ inducing the output state $\rho_{B^n}^c$. Further assume that $f(w) = f_1(w) \otimes f_2(w)$ for two functions $f_1: \intseq{1}{\mc} \to \calD\pr{\calH_A^{\proddist n_1}}$ and $f_2: \intseq{1}{\mc} \to \calD\pr{\calH_A^{\proddist n_2}}$ where $n = n_1 + n_2$. Let $\ket{\phi^w}_{RA^{n_1}}$ be a purification of $ \calM_{A\to A}^{\proddist n_1}(f_1(w))$ and $\sigma_{RB^{n_1}}^w \eqdef \id_R \otimes \calN_{A\to B}^{\proddist n_1} (\kb{\phi^w}_{RA^{n_1}})$. For any $\zeta > 0$, there exists an $(\mc, \ms, 2\zeta + 2\sqrt{\epsilon + \zeta} + 2\sqrt{\zeta + 2\sqrt{\epsilon + \zeta}} )^{\mathrm{CC-ES}}$ stego  protocol inducing the output state $\rho_{B^n}^s$ such that ${\norm{\rho_{B^n}^s - \rho_{B^n}^c}}_1 \leq \zeta$ provided that $\ms \leq  \min_{w\in \intseq{1}{\mc}}E_s(\ket{\phi}^w_{RB^{n_1}},  \ms^{\mathrm{CC}}(f_2, \zeta), \zeta )$.

 The stego protocol requires $\log \ms^{\mathrm{CC}}(f_2, \zeta) + \log K^{\mathrm{CC}}(f_2, \zeta)$ bits of shared key.
\end{theorem}
 \begin{remark}
 Our assumption that $f(w)$ decomposes as $f_1(w) \otimes f_2(w)$ for all $w$ holds for common codes for classical communication over quantum channels such as \cite{Wilde2013}.
 \end{remark}
We next show that if the cover protocol is an entanglement sharing code, there exists a stego protocol that  shares  both entanglement and classical randomness.
\begin{theorem}[entanglement sharing / classical randomness sharing]
\label{th:main-es} 
Let  the cover protocol be an $(\mc, \epsilon)^{\textnormal{ES}}$ code  $(\rho_{\tA A^n}, \calD_{B^n\to \tB })$ for $\calN^\pn_{A\to B}\circ \calM_{A\to A}^\pn$ inducing the output state $\rho_{B^n}^c$. If $\calH_A = \calH_B$ and $\calN_{A\to B} = \id_A$, for any $\zeta \geq 0$ and $\ms$, there exists an $(\mc, \ms, (\sqrt{\epsilon} + \zeta)^2 )^{\textnormal{ES-RS}} $\footnote{We have claimed the existence of an $(\mc, \ms, (\sqrt{2\epsilon - \epsilon^2 }+ \zeta )^{\textnormal{ES-RS}}$ stego protocol in \cite{Tahmasbi2019} because of an unfortunate mistake in our calculations. } stego protocol inducing the output state $\rho_{B^n}^c$ such that $\rho_{B^n}^s = \rho_{B^n}^s$ if
\begin{align}
\log \overline{M} \leq \sup_{a\in ]0, 1[} \pr{H^a((\id_{\tA}\otimes \calM_{A\to A}^\pn)(\rho_{\tA A^n})) -\frac{4}{a}\log \frac{2}{\zeta}}.
\end{align}
\end{theorem}
Finally we show that a cover protocol for quantum communication can be converted into a  quantum and classical communication stego protocol.
\begin{theorem} [quantum communication / classical communication]
\label{th:main-qc}
Let the cover protocol be an $(\mc, \epsilon)^{\textnormal{QC}}_{\textnormal{R}}$ code $(\calE_{W\to A^n}, \calD_{B^n\to W })$ inducing the output state $\rho_{B^n}^c$. Suppose that $\calE_{W\to A^n} = V_{W\to A^n} \rho_W V_{W\to A^n}^\dagger$ where $V_{W\to A^n}$ is an isometry. If $\calH_A = \calH_B$ and $\calN_{A\to B} = \id_A$, for all $\zeta > 4\sqrt{\epsilon}$, there exists an $(\mc, \ms, \max(\zeta, \epsilon))^{\textnormal{QC-CC}}_{\textnormal{R}}$ stego protocol inducing the output state $\rho_{B^n}^s$ such that ${\norm{\rho_{B^n}^s - \rho_{B^n}^c}}_1 \leq 2\epsilon + \zeta
$,\footnote{Note that $\rho_{B^n}^s$ and $\rho_{B^n}^c$ depend on $\rho_W$, and this inequality should hold for all choices of $\rho_W$.} provided that 
\begin{align}
\log \ms \leq \sup_{a\in]0, 1[}H^a\pr{{\calM_{A\to A}^c}^\pn\pr{\calE_{W\to A^n}\pr{\frac{1}{\ms} \one_W} }} 
- \frac{4}{a} \log \frac{2}{\zeta/2 - 2\sqrt{\epsilon}},
\end{align}
 where $\calM^c_{A\to A}$ is the complementary channel of $\calM_{A\to A}$.
\end{theorem}


\section{Examples}
\label{sec:example}
\subsection{Classical Codes with Product Structure}
\begin{definition}
\label{def:prod-encoder}
Let $k$ and $\ell$ be positive integers, and $\rho_{A^k}^1, \cdots, \rho_{A^k}^\ell\in \calD(\calH_A^{\proddist k})$. We say that an encoder $f:\intseq{1}{M} \to \calD(\calH_A^\pn)$ has \emph{a product structure with respect to} $\calP_{k, \ell} \eqdef \{\rho_{A^k}^1, \cdots, \rho_{A^k}^\ell\}$, if $n$ is divisible by $k$ and for all $w\in \intseq{1}{M}$, we have $f(w) = \otimes_{i=1}^{n/k} \sigma^i$ where $\sigma^1, \cdots, \sigma^{n/k} \in \calP_{k, \ell}$.
\end{definition}
\begin{remark}
Definition~\ref{def:prod-encoder} is useful when $n/k\gg 1$. Several explicit constructions of classical codes for quantum channels  are in this regime \cite{Wilde2013}. Moreover, from the standard random coding arguments, codes with large $n/k$  achieve the classical capacity of any quantum channel.
\end{remark}
Considering the cover classical communication code described in Theorem~\ref{th:main-cc}, we simplify the expressions for the rate  of the cypher message provided that the cover code has a product structure and $n/k$ is large enough. Let $\delta > 0$ and let the classical communication code have \emph{a product structure with respect to} $\calP_{k, \ell}$. There exist an integer $m$ depending on $\calP_{k, \ell}$, $\zeta, \delta > 0$ such that if $n/k \geq m$ the following two propositions hold.
\begin{proposition}
 For a noiseless channel, the number of bits of the cypher message is at least $\frac{n}{k}\pr{\min_{\rho \in \calP_{k, \ell}} H(\calM_{A\to A}^{\proddist k}(\rho))  - \delta}$. For a noisy channel, the number of bits of  the cypher message is at least 
$ \frac{n}{k}\pr{\inf_{\rho\in \calP_{k, \ell}}\sup_{\sigma_{XB^k}: \textnormal{tr}_X(\sigma_{XB^k}) = \rho} \avgI{X;B^k}_{\sigma} - \delta}$, using a shared secret key of $\delta n$ bits.
\end{proposition}
\begin{proposition}
 For a noiseless channel, the number of entangled qubits that the stego protocol of Theorem~\ref{th:main-cc-es} would generate is at least $\frac{n}{k}\pr{\min_{\rho \in \calP_{k, \ell}} H(\calM_{A\to A}^{\proddist k}(\rho))  - \delta}$.
The required number of shared secret key bits is $O(\log n)$.
\end{proposition}
\subsection{Gaussian States}
Although we have assumed so far that all Hilbert spaces are finite dimensional, the proof of the first part of Theorem~\ref{th:main-cc} carries over to infinite dimensional spaces  since the leftover hash lemma still holds for such a setting. Gaussian channels form an important class of infinite dimensional channels, which models optical channels. Let  $A$ and $B$ be single mode  bosonic systems, $\calN_{A\to B}$ be noiseless, $\calM_{A\to A}$ be a Gaussian channel, and $f(w)$ be a Gaussian state for all $w$. Denoting the symplectic spectra of $\rho_{B^n}$ by $(\nu_1^w, \cdots, \nu_n^w)$, we have \cite[Eq. (108)]{Adesso2014}
$H^a(\rho_{B^n}^w) = -\frac{\sum_{i=1}^n \log\pr{\eta_{1+a}(\nu_i^w)}}{a}$, where
$\eta_\alpha(x) \eqdef 2^\alpha/((x+1)^\alpha - (x-1)^\alpha)$. The number of bits of the cypher message would then be 
\begin{align}
\log \ms = \min_{\wc \in \intseq{1}{\mc}} \sup_{a\in ]0, 1[} -\frac{\sum_{i=1}^n \log\pr{\eta_{1+a}(\nu_i^w)}}{a} -\frac{4}{a} \log \frac{2}{\zeta}.
\end{align}

\begin{figure}[t]
  \centering
  \includegraphics[width=.6\linewidth]{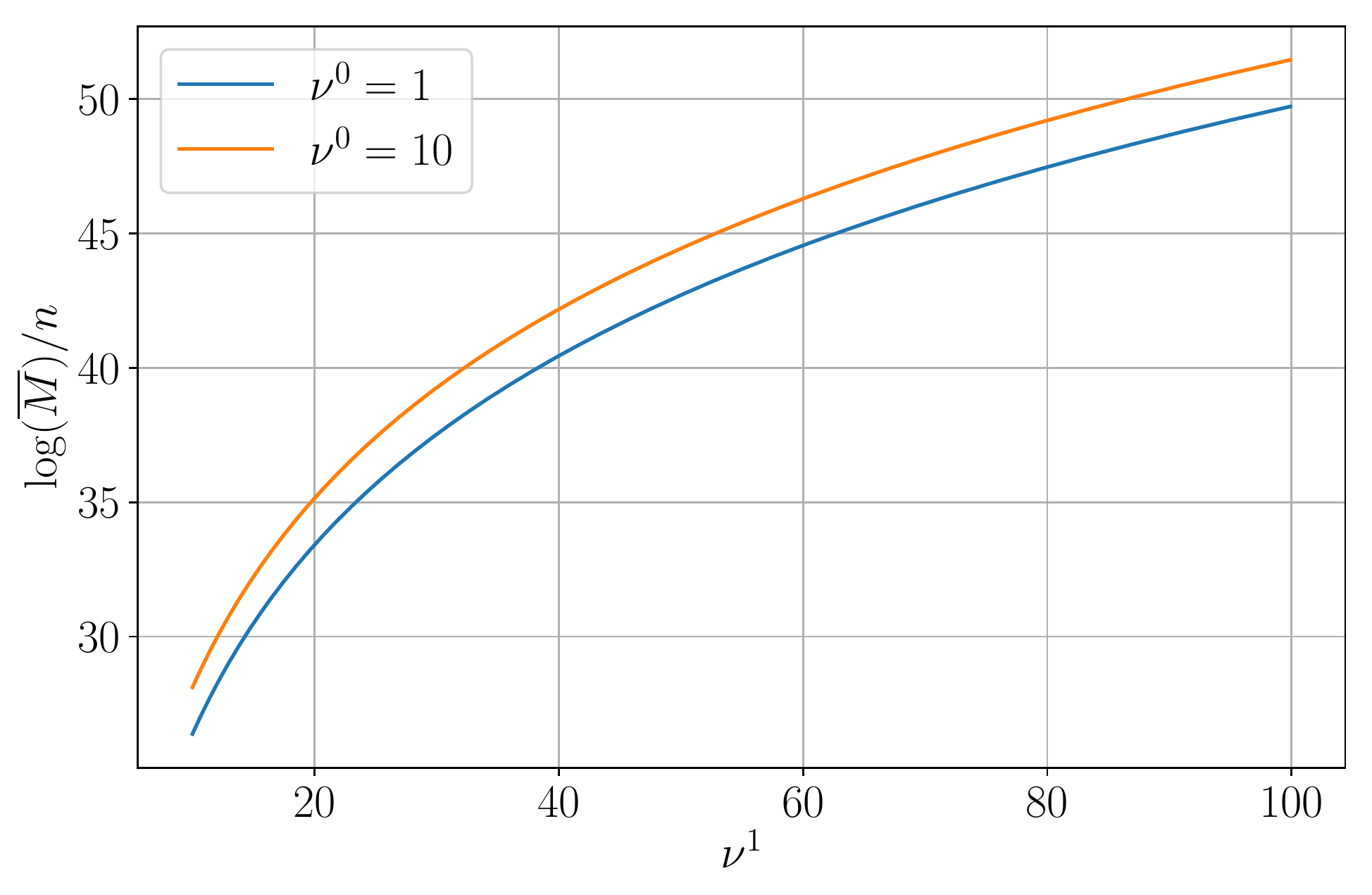}
  \caption{Rate of the cypher message vs the symplectic eigenvalues of $\calM_{A\to A}(\rho_A^0)$ and $\calM_{A\to A}(\rho_A^1)$, $\nu^0$ and $\nu^1$}
    \label{fig:example}

\end{figure}

We now suppose that the cover code uses a binary modulation, i.e., for two states $\rho_{A}^0$ and $\rho_{A}^1$, we have $f(w) = \otimes_{i=1}^n \rho_{A}^{x_{w, i}}$ for all $\wc\in \intseq{1}{\mc}$. Let $\nu^0$ and $\nu^1$ be the symplectic eigenvalue of $\calM_{A\to A}(\rho_A^0)$ and $\calM_{A\to A}(\rho_A^1)$, respectively, with $\nu^0\leq \nu^1$. Upon defining $r \eqdef \min_{w\in \intseq{1}{\mc}} \sum_{i=1}^n x_{w, i}$, we have
\begin{align}
\log \ms =  \sup_{a\in ]0, 1[} -\frac{ (n-r) \log\pr{\eta_{1+a}(\nu^0)} + r \log \pr{\eta_{1+a}(\nu^1)}}{a} -\frac{4}{a} \log \frac{2}{\zeta}.
\end{align}
We plot the rate of the cypher message for $n=10^6$, $r=n/2$, $\zeta = 10^{-3}$ in Fig.~\ref{fig:example}.

\subsection{Quantum Codes of \cite{Sutherland2018a}}
Consider a Kraus representation $\{F_j\}_{j\in \calJ}$ of $\calM_{A\to A}$ such that $\tr{F_j^\dagger F_{j'}} = \indic{j=j'} d_j$. This defines a Kraus representation $\{\mathbf{F}_{\mathbf{j}}\}_{\mathbf{j} \in \calJ^n}$ for $\calM_{A\to A}^\pn$, where $\mathbf{F}_{\mathbf{j}} \eqdef F_{j_1}\otimes \cdots \otimes F_{j_n}$. Let $\calT_{\epsilon}$ be the typical subset of $\mathbf{F}_{\mathbf{j}} \eqdef F_{j_1}\otimes \cdots \otimes F_{j_n}$ as defined in \cite{Klesse2007}. If $\Pi$ is the projector onto the sub-space of inputs defined by the code, we assume that for all $\mathbf{F}_{\mathbf{j}} \in \calT_\epsilon$, we have $ \mathbf{F}_{\mathbf{j}} \Pi = p_{\mathbf{j}} \mathbf{U}_{\mathbf{j}} \Pi$, where $p_{\mathbf{j}} = p_{j_1} \times \cdots \times p_{j_n}$ for a probability distribution $\{p_j\}$ on $\calJ$, and $\mathbf{U}_{\mathbf{j}} = U_{j_1} \otimes \cdots \otimes U_{j_n}$ for unitaries $\{U_j\}$ on $\calH_A$. 
\begin{proposition}
For all $\delta>0$ and $n$ large enough, we have
\begin{align}
\sup_{a\in]0, 1[}H^a\pr{{\calM_{A\to A}^c}^\pn\pr{\calE_{W\to A^n}\pr{\frac{1}{\ms} \one_W} }} \geq n\pr{-\sum_j p_j \log p_j-\delta}. 
\end{align}
\end{proposition}
\subsection{Random Quantum Codes}
\begin{proposition}
Let $\calS$ be a random  $M$-dimensional subspace of $\calH_A^\pn$ distributed according to the Haar measure, and $\Pi$ denote the projector onto $\calS$. Let $\calM_{A\to A}$ be a quantum channel with an isometric extension $V_{A\to AE}$. For all $\delta> 0$, there exists $n$ large enough such that
\begin{align}
\E[\calS]{\sup_{a\in]0, 1[}H^a\pr{{\calM_{A\to A}^c}^\pn\pr{\frac{\Pi}{M}}}} \geq n \pr{ \log \min(\textnormal{rank}\pr{ \textnormal{tr}_A(VV^\dagger)}, \textnormal{rank} \pr{\textnormal{tr}_E(VV^\dagger))} -\delta}
\end{align}
\end{proposition}
\begin{IEEEproof}
Let $\ket{g^1}, \cdots, \ket{g^M}$ be $M$ random independent Gaussian vectors  in $\calH_A^\pn$ as defined in \cite{Hayden2008a}.  Since the distribution of $\textnormal{span}(\ket{g^1}, \cdots, \ket{g^M})$ is the same as the distribution of $\calS$ by \cite{Hayden2008a}, we take $\calS = \textnormal{span}(\ket{g^1}, \cdots, \ket{g^M})$. Defining $G \eqdef \sum_{j=1}^M \ket{g^j}\bra{g^j}$, the vectors $\{\ket{\phi^j}_{A^n} \eqdef G^{-\frac{1}{2}} \ket{g^j}\}_{j\in \intseq{1}{M}}$ form an orthonormal basis for $\calS$. One can check that $\ket{\psi^j}_{A^nE^n} \eqdef V_{A\to AE} \ket{\phi^j}$ has a uniform distribution over all unit vectors in $(\textnormal{range} VV^\dagger)^\pn$. Therefore, we have
\begin{align}
\E[\calS]{H^a\pr{{\calM_{A\to A}^c}^\pn\pr{\frac{\Pi}{M}}} }
&\stackrel{(a)}{=}\E[\calS]{ H^a\pr{\frac{1}{M}\sum_{j=1}^M {\calM_{A\to A}^c}^\pn\pr{\ket{\phi^j}\bra{\phi^j}} }}\\
&\geq\frac{1}{M}\sum_{j=1}^M \E[\calS]{ H^a\pr{ {\calM_{A\to A}^c}^\pn\pr{\ket{\phi^j}\bra{\phi^j}} }}\\
&=  \frac{1}{M}\sum_{j=1}^M \E[\calS]{H^a\pr{ { \psi^j_{E^n} }}}\\
&\stackrel{(b)}{\geq} n \log \textnormal{rank}\pr{ \textnormal{tr}_A(VV^\dagger)} + n \log \textnormal{rank}\pr{ \textnormal{tr}_E(VV^\dagger)} \nonumber\\
&- 2 \log\pr{\sqrt{\textnormal{rank}\pr{ \textnormal{tr}_A(VV^\dagger)}^n} + \sqrt{\textnormal{rank}\pr{ \textnormal{tr}_E(VV^\dagger)} ^n}}\\
&\geq n \log \min(\textnormal{rank}\pr{ \textnormal{tr}_A(VV^\dagger)}, \textnormal{rank} \pr{\textnormal{tr}_E(VV^\dagger))}- \log 2, 
\end{align}
where $(a)$ follows from the concavity of R\'enyi entropy, and $(b)$ follows from the bound in the proof of \cite[Lemma III.1]{Hayden2008}.
\end{IEEEproof}
\section{Proofs}
\label{sec:proof}
\subsection{One-shot Results}
In this section, we develop  one-shot coding results stated in terms of R\'enyi mutual information. We shall specialize them  to prove our main results in Section~\ref{sec:main-proof}. We first derive an achievability result stating that there exists a classical communication code for a cq channel inducing a pre-specified state at the output. Our proof is based on combining quantum channel coding and channel resolvability results.
\begin{lemma}
\label{lm:res-rel-one}
Let $\rho_{XB} = \sum_{x\in\calX} P_X(x) \ket{x}\bra{x} \otimes \rho_B^x$ be a cq state. Let $M$ and $K$ be positive integers. For each $s\in\intseq{1}{K}$, there exist an encoding function $g_s:\intseq{1}{M}\to\calX$ and a POVM $\mathbf{\Gamma}_s = \{\Gamma_s^w\}_{w\in\intseq{1}{M}}$ such that 
\begin{align}
\frac{1}{MK}\sum_{s=1}^K\sum_{w=1}^M \tr{\Gamma_s^w \rho_B^{g_s(w)}} \geq 1-\epsilon, \text{ and }
\label{eq:res}
{\norm{\frac{1}{MK}\sum_{s=1}^K\sum_{w=1}^M \rho_B^{g_s(w)} - \rho_B}}_1 \leq \delta,
\end{align}
provided that  
\begin{align}
\log M \leq \sup_{a\in]0, 1[} \left[\ih{a}{X;B}{\rho} -\frac{1}{a}\log \nu(\rho_X\otimes \rho_B) - \frac{4}{a}\log \frac{12}{\epsilon}\right],
\end{align}
and
\begin{align}
\log MK \geq \inf_{a<0} \left[\ir{a}{X;B}{\rho} + \log \nu(\rho_B) +  \pr{2-\frac{2}{a}} \log \frac{12}{\delta} \right]  .
\end{align}
\end{lemma}
\begin{IEEEproof}
We consider $K$ independently generated random encoders $G_s:\intseq{1}{M}\to\calX$ where $G_s(1), \cdots, G_s(M)$ are \ac{iid} according to $P_X$. By \cite[Theorem 1]{Wang2012}, for all $\epsilon > 0$, there exists a POVM  such that
 \begin{align}
 \E[G_s]{\frac{1}{M}\sum_{w=1}^M \tr{\Gamma_s^w \rho_B^{G_s(w)}}} \geq 1-\epsilon,
 \end{align}
 if $\log M \leq D_H^{\epsilon/2}(\rho_{XB}\| \rho_X \otimes \rho_B) - \log \frac{1}{\epsilon} - 4$ where $ D_H^\epsilon(\rho\|\sigma) \eqdef -\log \inf_{Q:0\leq Q \leq \one, \tr{Q\rho} \geq 1-\epsilon} \tr{Q\sigma}$. By \cite[Theorem 2]{Ogawa2004}, for all $\gamma >0$, there exists   an operator $0\leq Q\leq \one_{XB}$ such that  for all $a\in]0, 1[$,
 \begin{align}
 \tr{Q (\rho_X\otimes \rho_B)} &\leq\nu(\rho_X \otimes \rho_B) e^{-\ih{a}{X;B}{\rho}a - \gamma (1-a)}\\
  \tr{(\one_{XB } - Q) \rho_{XB}} &\leq  \nu(\rho_X \otimes \rho_B) e^{-\ih{a}{X;B}{\rho} a+ \gamma a}.
 \end{align}
Choosing $\gamma = \ih{a}{X;B}{\rho}  -a^{-1}(\log \nu(\rho_X \otimes \rho_B) + \log 2/\epsilon )$ yields that
 \begin{align}
 D_H^{\epsilon/2}(\rho_{XB}\| \rho_X \otimes \rho_B) - \log \frac{1}{\epsilon} - 4 
 &\geq \ih{a}{X;B}{\rho} -\frac{1}{a}\log \nu(\rho_X\otimes \rho_B) - \frac{1}{a}\log \frac{2}{\epsilon} - \log \frac{1}{\epsilon} - 4\nonumber\\
 &\geq  \ih{a}{X;B}{\rho} -\frac{1}{a}\log \nu(\rho_X\otimes \rho_B) - \frac{4}{a}\log \frac{4}{\epsilon}.
 \end{align}
 To obtain \eqref{eq:res}, note that by \cite[Lemma~9.2]{Hayashi2006}, we have for all $a< 0$
 \begin{align}
 \E[G_1, \cdots, G_K]{{\norm{\frac{1}{MK}\sum_{s=1}^K\sum_{w=1}^M \rho_B^{G_s(w)} - \rho_B}}_1 } 
 &\leq \inf_{\lambda > 0} 2\sqrt{e^{a(\log \lambda- \ir{a}{X;B}{\rho})}} + \sqrt{\frac{\lambda \nu(\rho_B)}{MK}}.
 \end{align}
 Choosing $\log \lambda = \ir{a}{X;B}{\rho} + \frac{2}{a} \log \delta/4$ and $\log MK \geq -2\log \frac{\delta}{2} +\log \lambda +\log \nu(\rho_B)$, we obtain that 
 \begin{align}
  \E[G_1, \cdots, G_K]{{\norm{\frac{1}{MK}\sum_{s=1}^K\sum_{w=1}^M \rho_B^{G_s(w)} - \rho_B}}_1 } 
 &\leq\delta.
\end{align}  
Finally,  Markov's inequality and the bounds on the expected values imply the existence of the desired code.
\end{IEEEproof}
We now prove the existence of a code for transmission of a classical message over a \emph{noiseless} classical  channel while a pre-specified distribution is induced at the output of the channel. We show that no key is required in this case. The idea of the proof is similar to  \cite[Lemma 2]{Renes2011}.
\begin{lemma}
\label{lm:clas-hash}
Let $P_X$  be a \ac{PMF} over  $\calX$, and $Q_W$ be uniform distribution  over $\intseq{1}{M}$ for $M\in \N^+$. Let $(W, X, \widehat{W})$ be distributed according to $Q_{WX \widehat{W}}(w, {x}, \widehat{w}) \eqdef Q_W(w)Q_{{X}|W}(x| w) \indic{f({x}) = \widehat{w}}$ for a conditional \ac{PMF} $Q_{X|W}$ and a function $f:\calX \to \intseq{1}{M}$.  For all $\epsilon > 0$, there exists $Q_{{X}|W}$ and $f$ such that 
\begin{align}
\label{eq:clas-hash1}{\norm{Q_{{X}} - P_X}}_1 &\leq \epsilon ,\\
\label{eq:clas-hash2}\P[Q]{W \neq \widehat{W}} &\leq  \epsilon,
 \end{align}
provided that $\log M \leq \sup_{a\in]0, 1[}H^a(P_X)  - \frac{4}{a} \log \frac{2}{\epsilon}$.
\end{lemma}
\begin{IEEEproof}
Let  $P_{WX \widehat{W}}$ be another distribution for $({W}, X, \widehat{{W}})$ defined as $$P_{{W}X \widehat{{W}}} ({w}, x, \widehat{{w}})= P_X(x)\indic{g(x) = {w}, g(x) = \widehat{{w}}}$$ for a function $g:\calX \to \intseq{1}{M}$. Using a privacy amplification result \cite[Corollary 5.6.1]{renner2008security} and a bound on smooth min-entropy in terms of R\'enyi entropy  \cite[Theorem 7]{Tomamichel2009}, there exists $g$ such that  ${\norm{P_{W} - Q_W}}_1 \leq \epsilon$ when $\log M \leq \sup_{a\in]0, 1[} H^a(P_X)  - \frac{4}{a} \log \frac{2}{\epsilon} $. It is enough to show that \eqref{eq:clas-hash1} and \eqref{eq:clas-hash2} hold for $f \eqdef g$ and 
\begin{align}
Q_{X|W}(x|w)  &\eqdef \begin{cases}\frac{P_{XW}(x,w)}{P_W(w)}&\quad P_W(w)\neq 0\\ P_X(x)&\quad P_W(w) = 0\end{cases}
\end{align}

 Note that $P_{\widehat{W}|XW}(\widehat{w}|x,w) = \indic{f(x) = \widehat{w}}$ and we have $P_{XW}(x, w) = P_W(w) Q_{X|W}(x|w)$  for all $w, x$. We thus have
\begin{align}
{\norm{Q_{WX\widehat{W}} - P_{WX\widehat{W}}}}_1
&= \sum_{x, w, \widehat{w}} | Q_W(w)Q_{{X}|W}({x}| w) \indic{f({x}) = \widehat{w}} - P_W(w) Q_{X|W}(x|w) \indic{f({x}) = \widehat{w}} | \\
&\leq \sum_w |Q_W(w) - P_W(w)| = {\norm{Q_W - P_W}}_1\leq \epsilon.
\end{align}
By the data processing inequality, \eqref{eq:clas-hash1} holds. Since for any two distributions $P$ and $Q$, we have ${\norm{P - Q}}_1 = 2\sup_{\calA} P(\calA) - Q(\calA)$, we have
\begin{align}
\P[Q]{W \neq \widehat{W}} 
& \leq \P[P]{W \neq \widehat{W}}+ \frac{1}{2}{\norm{Q_{WX\widehat{W}} - P_{WX\widehat{W}}}} \\
&= \frac{1}{2}{\norm{Q_{WX\widehat{W}} - P_{WX\widehat{W}}}} \leq \epsilon.
\end{align}
\end{IEEEproof}
We extend Lemma~\ref{lm:clas-hash} to the quantum setting in the following corollary.
\begin{corollary}
\label{cor:quant-hash}
Let $M\in \mathbb{N}^+$, $\epsilon>0$, $\calH$ be a finite dimensional Hilbert space, and $\rho$ be a density operator on $\calH$.  Suppose that  $\log M \leq \sup_{a\in ]0, 1[} H^a(\rho) -\frac{4}{a} \log \frac{2}{\epsilon}.
$ There exist a function $g:\intseq{1}{M} \to \calD(\calH)$ and a POVM $\mathbf{\Lambda} =\{\Lambda^w\}_{w\in\intseq{1}{M}}$ such that
\begin{align}
\label{eq:quant-hash1}{\norm{ \frac{1}{M}\sum_{w=1}^M g(w) - \rho}}_1 &\leq \epsilon,\displaybreak[0]\\
\label{eq:quant-hash2}\frac{1}{M} \sum_{w=1}^M \tr{\Lambda^w g(w)}&\geq 1- \epsilon.\displaybreak[0]
\end{align}
\end{corollary}
\begin{IEEEproof}
Considering an eigen-decomposition of $\rho$ as $\sum_{i=1}^d P_X(x_i) \ket{x_i}\bra{x_i}$ and defining $\calX \eqdef \{x_1, \cdots, x_d\}$, we apply Lemma~\ref{lm:clas-hash} to $P_X$ to obtain a conditional \ac{PMF} $Q_{X|W}$ and a function $f:\calX\to \intseq{1}{M}$ satisfying \eqref{eq:clas-hash1} and \eqref{eq:clas-hash2}. Let $Q_{WX\widehat{W}}$ be as defined in Lemma~\ref{lm:clas-hash}. We then define
\begin{align}
\label{eq:gw-def}
g(w) &\eqdef \sum_{x\in \calX} Q_{X|W}(x|w) \ket{x}\bra{x},\\
\label{eq:lambdaw-def}
\Lambda^w &\eqdef \sum_{x: f(x)= w} \ket{x}\bra{x}.
\end{align}
Substituting \eqref{eq:gw-def} in \eqref{eq:quant-hash1}, we obtain
\begin{align}
{\norm{\frac{1}{M}\sum_{w=1}^M g(w) - \rho}}_1 
&= {\norm{\frac{1}{M}\sum_{w=1}^M  \sum_{x\in \calX} Q_{X|W}(x|w) \ket{x}\bra{x}- \rho}}_1\displaybreak[0]\\
&= {\norm{\sum_{x\in \calX}\sum_{w=1}^M Q_W(w) Q_{X|W}(x|w) \ket{x}\bra{x} - \rho}}_1\displaybreak[0]\\
&= {\norm{\sum_{x\in \calX} Q_X(x) \ket{x} \bra{x} -\sum_{x\in \calX} P_X(x) \ket{x} \bra{x} }}_1\displaybreak[0]\\
&= {\norm{Q_X - P_X}}_1\leq \epsilon.
\end{align}
Moreover,
\begin{align}
\frac{1}{M}\sum_{w=1}^M\tr{\Lambda^w g(w)}
&= \frac{1}{M}\sum_{w=1}^M\tr{\pr{ \sum_{x: f(x)= w} \ket{x}\bra{x}} \pr{\sum_{x\in \calX} Q_{X|W}(x|w) \ket{x}\bra{x}}}\displaybreak[0]\\
&= \frac{1}{M}\sum_{w=1}^M\tr{{ \sum_{x: f(x)= w} Q_{X|W}(x|w) \ket{x}\bra{x}}}\displaybreak[0]\\
&= \sum_{x, w} Q_W(w) Q_{X|W}(x|w) \indic{f(x) = w}\displaybreak[0]\\
&=  \sum_{x, w, \widehat{w}} Q_W(w) Q_{X|W}(x|w) \indic{f(x) = \widehat{w}} \indic{w = \widehat{w}}\\
&= \P[Q]{W = \widehat{W}} \leq \epsilon.
\end{align}
\end{IEEEproof} 
%
%

\subsection{Proof of Main Results}
\label{sec:main-proof}
\begin{IEEEproof}[Proof of Theorem~\ref{th:main-cc}]
We separately prove the two parts of the theorem. Let the code $(f, \mathbf{\Lambda})$ be the cover protocol, and the main channel be noiseless. By Corollary~\ref{cor:quant-hash},  for every $\zeta \geq 0$, provided that 
\begin{align}
\log \ms \leq \inf_{\wc} \sup_{a\in ]0, 1[} H^a(\calM_{A\to A}^\pn(f(w))) -\frac{4}{a} \log \frac{2}{\epsilon},
\end{align}
 there exist a function $g_w:\intseq{1}{\ms}\to \calD\pr{\calH_A^\pn}$ and a  POVM $\mathbf{\Gamma}_w = \{{\Gamma}^{\ws}_\wc\}_{\ws\in\intseq{1}{\ms}}$ such that
\begin{align}
{\norm{ \frac{1}{\ms}\sum_{\ws=1}^{\ms} g_\wc(\ws) - \calM_{A\to A}^\pn(f(\wc))}}_1 \leq \zeta, \text{ and }
\frac{1}{\ms} \sum_{\ws=1}^\ms \tr{\Gamma^\ws_\wc g_\wc(\ws)}\geq1-\zeta. \label{eq:indv-steg-pe-cc}
\end{align}
We define the stego protocol as follows. Let $\overline{f}:\intseq{1}{\mc\ms}\to \calD\pr{\calH_{A}^\pn}$ be defined as  $\overline{f}(\wc(\ms-1)+\ws) \eqdef g_\wc(\ws)$. We define a POVM $  \{\sqrt{{\Lambda}^\wc}\Gamma_\wc^\ws \sqrt{{\Lambda}^\wc}\}_{w\in \intseq{1}{M},\ws\in \intseq{1}{\ms} }$, which is equivalent to first measuring $\mathbf{\Lambda}$ and then measuring $\mathbf{\Gamma}_{\wc}$. This is a valid POVM, since every $\sqrt{{\Lambda}^\wc}\Gamma_\wc^\ws \sqrt{{\Lambda}^\wc}$ is a positive operator and $
\sum_{\wc\ws}\sqrt{{\Lambda}^\wc}\Gamma_\wc^\ws \sqrt{{\Lambda}^\wc}
\stackrel{(a)}{=} \sum_\wc \Lambda^w
\stackrel{(b)}{=} \one_B,$
where $(a)$ follows since $\mathbf{\Gamma}_\wc$ is a valid POVM, and $(b)$ follows since $\mathbf{\Lambda}$ is a valid POVM. Note next that 
\begin{align}
{\norm{\rho_{B^n}^c - \rho_{B^n}^s}}_1
&= {\norm{\frac{1}{\mc} \sum_{\wc = 1}^\mc\calM_{A\to A}^\pn(f(\wc)) - \frac{1}{\mc\ms} \sum_{\wc\ws} g_\wc(\ws)}}_1\\
&\stackrel{(a)}{\leq} \frac{1}{\mc} \sum_{\wc}^\mc {\norm{\calM_{A\to A}^\pn(f(\wc))- \frac{1}{\ms} \sum_\ws g_\wc(\ws)}}_1 \leq \zeta,
\end{align}
where $(a)$ follows from the convexity of the trace norm.
The probability of correct decoding is also
\begin{align}
 \label{eq:pe-first-cc-noiseless}\frac{1}{\mc\ms} \sum_{\wc\ws} \tr{\sqrt{{\Lambda}^\wc} \Gamma_\wc^\ws  \sqrt{{\Lambda}^\wc}g_\wc(\ws)}\displaybreak[0]
&= \frac{1}{\mc\ms} \sum_{\wc\ws} \tr{\Gamma_\wc^\ws  \sqrt{{\Lambda}^\wc}g_\wc(\ws)\sqrt{{\Lambda}^\wc} }\displaybreak[0]\\
&= \frac{1}{\mc\ms} \sum_{\wc\ws} \tr{\Gamma_\wc^\ws  g_\wc(\ws)}\nonumber \\
&+  \frac{1}{\mc\ms} \sum_{\wc\ws} \tr{\Gamma_\wc^\ws\pr{  \sqrt{{\Lambda}^\wc}g_\wc(\ws)\sqrt{{\Lambda}^\wc}-   g_\wc(\ws)} }.\label{eq:pe-cc-noiseless}
\end{align}
 We also have $ \frac{1}{\mc\ms} \sum_{\wc\ws} \tr{\Gamma_\wc^\ws  g_\wc(\ws)} \geq 1-\zeta$ by \eqref{eq:indv-steg-pe-cc}. To lower-bound the second term in \eqref{eq:pe-cc-noiseless}, we have
\begin{align}
\frac{1}{\mc\ms} \sum_{\wc\ws} \tr{\Gamma_\wc^\ws\pr{  \sqrt{{\Lambda}^\wc}g_\wc(\ws)\sqrt{{\Lambda}^\wc}-   g_\wc(\ws)} } 
& \geq - \frac{1}{\mc\ms} \sum_{\wc\ws}  {\norm{\Gamma_\wc^\ws}}_\infty {\norm{ \sqrt{{\Lambda}^\wc}g_\wc(\ws)\sqrt{{\Lambda}^\wc}-   g_\wc(\ws)}}_1\displaybreak[0]\\
&\geq- \frac{1}{\mc\ms} \sum_{\wc\ws} {\norm{ \sqrt{{\Lambda}^\wc}g_\wc(\ws)\sqrt{{\Lambda}^\wc}-   g_\wc(\ws)}}_1\displaybreak[0]\\
&\stackrel{(a)}{\geq} - \frac{1}{\mc\ms} \sum_{\wc\ws} 2\sqrt{1-\tr{{\Lambda}^\wc g_\wc(\ws)}}\displaybreak[0]\\
&\stackrel{(b)}{\geq} -2\sqrt{1- \frac{1}{\mc} \sum_{\wc} \tr{{\Lambda}^\wc\pr{ \frac{1}{\ms} \sum_\ws g_\wc(\ws)}}},
\end{align}
where $(a)$ follows from the gentle operator lemma \cite{Winter1999}, and $(b)$ follows from Jensen's inequality and the concavity of $x\mapsto \sqrt{1-x}$.
We also lower-bound
\begin{align}
 &\frac{1}{\mc} \sum_{\wc} \tr{{\Lambda}^\wc\pr{ \frac{1}{\ms} \sum_\ws g_\wc(\ws)}} \displaybreak[0]\\
 &~~~=  \frac{1}{\mc} \sum_{\wc} \tr{{\Lambda}^\wc \calM_{A\to A}^\pn(f(w))} +  \frac{1}{\mc} \sum_{\wc} \tr{{\Lambda}^\wc\pr{ \frac{1}{\ms} \sum_\ws g_\wc(\ws) - \calM_{A\to A}^\pn(f(w))}}\displaybreak[0]\\
  &~~~\geq 1 - \epsilon +  \frac{1}{\mc} \sum_{\wc} \tr{{\Lambda}^\wc\pr{ \frac{1}{\ms} \sum_\ws g_\wc(\ws) - \calM_{A\to A}^\pn(f(w))}}\displaybreak[0]\\
  &~~~\geq 1-\epsilon  - \frac{1}{\mc} \sum_{\wc} {\norm{\Lambda^\wc}}_\infty {\norm{\frac{1}{\ms} \sum_\ws g_\wc(\ws) - \calM_{A\to A}^\pn(f(w))}}_1\displaybreak[0]\\
  &~~~\geq 1-\epsilon  - \frac{1}{\mc} \sum_{\wc}  {\norm{\frac{1}{\ms} \sum_\ws g_\wc(\ws) - \calM_{A\to A}^\pn(f(w))}}_1\displaybreak[0]\geq 1-\epsilon-\zeta.\label{eq:pe-last-cc-noiseless}
\end{align}
Let $\zeta$, $\xi$, $\rho_{B^n}^\wc$, $\ms_\wc$, $\ms$, and $\ks$ be defined as in the statement of Theorem~\ref{th:main-cc}, and the main channel be noisy. We assume without loss of generality that $\ms_w$ is divisible by $\ms$ for all $w$, otherwise we define $\ms_\wc' \eqdef \lfloor \ms_\wc/\ms\rfloor \ms$, and $|\log \ms_\wc - \log \ms_\wc'| \leq 1$.  By Lemma~\ref{lm:res-rel-one}, for each $\wc$, there exist $\ks$ encoding functions $g_{s, \wc}: \intseq{1}{\ms_\wc} \to \calD(\calH_A^\pn)$ and $\ks$ POVMs $\mathbf{\Gamma}_{s, \wc} = \{\Gamma_{s, \wc}^\ws\}_{\ws\in \intseq{1}{\ms_\wc}}$ such that $\frac{1}{\ms_\wc K}\sum_{s\ws} \tr{\Gamma_{s, \wc}^\ws \rho_{B^n}^\wc } \geq 1-\zeta$, and
\begin{align}
{\norm{\frac{1}{\ms_\wc K}\sum_{s\ws}\calN_{A\to B}^\pn(g_{s, \wc}(\ws)) - \rho_{B^n}^\wc}}_1 \leq \xi.
\end{align}

We define the stego protocol as follows. For $s\in \intseq{1}{K}$, $\wc\in\intseq{1}{\mc}$, and $\ws\in \intseq{1}{\ms}$, we define $\mu_\wc \eqdef \frac{\ms_\wc}{\ms}$,  $\overline{f}_s((\wc-1)\ms + \ws) \eqdef \frac{1}{\mu_\wc}\sum_{i=1}^{\mu_\wc} g_{s, \wc}((\ws-1)\mu_\wc + i)$, and $
\overline{\Lambda}_s^{(\wc-1)\ms + \ws} \eqdef \sum_{j=1}^{\mu_\wc} \sqrt{\Lambda^\wc} \Gamma_{s, \wc}^{(\ws-1)\mu_\wc + j} \sqrt{\Lambda^\wc}$.
As done previously, one can show that for each $s\in \intseq{1}{K}$, $\overline{\mathbf{\Lambda}}_s = \{\overline{\Lambda}_s\}$ is a valid POVM. Note also that
\begin{align}
&\frac{1}{K\mc\ms} \sum_{s, \wc, \ws} \tr{\Lambda_s^{(\wc-1)\ms + \ws} f_s((\wc-1)\ms + \ws)}\displaybreak[0]\\
&\phantom{===}= \frac{1}{K\mc\ms} \sum_{s, \wc, \ws} \tr{\pr{\sum_{i=1}^{\mu_\wc} \sqrt{\Lambda^\wc} \Gamma_{s, \wc}^{(\ws-1)\mu_\wc + i} \sqrt{\Lambda^\wc}}\pr{\frac{1}{\mu_\wc}\sum_{j=1}^{\mu_\wc} g_{s, \wc}((\ws-1)\mu_\wc + j)}}\displaybreak[0]\\
&\phantom{===}\geq  \frac{1}{K\mc\ms} \sum_{s, \wc, \ws} \tr{\frac{1}{\mu_\wc} \sum_{i=1}^{\mu_\wc} \sqrt{\Lambda^\wc} \Gamma_{s, \wc}^{(\ws-1)\mu_\wc + i} \sqrt{\Lambda^\wc} g_{s, \wc}((\ws-1)\mu_\wc + i)}\displaybreak[0]\\
 &\phantom{===}\stackrel{(a)}{=} \frac{1}{K\mc} \sum_{s, \wc} \frac{1}{\ms_\wc}\sum_{\ws=1}^{\ms_\wc} \tr{ \sqrt{\Lambda^\wc} \Gamma_{s, \wc}^{\ws} \sqrt{\Lambda^\wc} g_{s, \wc}((\ws)},\displaybreak[0]
\end{align}
where $(a)$ follows since the index $(\ws - 1)\mu_\wc +i$ is changing from $1$ to $\ms_\wc$.
Repeating   calculations similar to \eqref{eq:pe-first-cc-noiseless}-\eqref{eq:pe-last-cc-noiseless}, we obtain that
\begin{align}
\frac{1}{K\mc\ms} \sum_{s, \wc, \ws} \tr{\Lambda_s^{(\wc-1)\ms + \ws} f_s((\wc-1)\ms + \ws)} \geq 1-\zeta -2\sqrt{\zeta + \epsilon}.
\end{align}
Furthermore, we have
\begin{align}
&{\norm{\frac{1}{\mc} \sum_{\wc=1}^\mc \rho_{B^n}^\wc - \frac{1}{\ks\mc\ms} \sum_{s, \wc\ws} \calN_{A\to B}^\pn(\overline{f}_s((\wc-1)\ms + \ws))}}_1\displaybreak[0]\\
&\phantom{========}\stackrel{(a)}{\leq} \frac{1}{\mc} \sum_{\wc=1}^\mc {\norm{ \rho_{B^n}^\wc  - \frac{1}{\ks\ms} \sum_{s\ws} \calN_{A\to B}^\pn(\overline{f}_s((\wc-1)\ms + \ws))}}_1\displaybreak[0]\\
&\phantom{========}=  \frac{1}{\mc} \sum_{\wc=1}^\mc {\norm{ \rho_{B^n}^\wc  - \frac{1}{\ks\ms} \sum_{s\ws} \calN_{A\to B}^\pn\pr{\frac{1}{\mu_\wc}\sum_{i=1}^{\mu_\wc} g_{s, \wc}((\ws-1)\mu_\wc + i)}}}_1\displaybreak[0]\\
&\phantom{========}=  \frac{1}{\mc} \sum_{\wc=1}^\mc {\norm{ \rho_{B^n}^\wc - \frac{1}{\ks\ms_\wc} \sum_{s\ws} \calN_{A\to B}^\pn\pr{ g_{s, \wc}(\ws)}}}_1\leq \xi,\displaybreak[0]
\end{align}
where $(a)$ follows from the convexity of the trace norm.

\end{IEEEproof}

\begin{IEEEproof}[Proof of Theorem~\ref{th:main-cc-es}] Intuitively, Alice splits the transmission into two part. Alice generates a purification of the state supposed to be transmitted in the first part,  keeps the reference system for herself, and transmits the state over the channel, which results in a shared entangled state between Alice and Bob. Alice and Bob use an entanglement distillation protocol to distill perfect entanglement in the second part of the transmission. This might require classical communication, which can be achieved by using the result of Theorem~\ref{th:main-cc}.
To formally state our protocol, we first need a generalization of the gentle measurement lemma.
\begin{proposition}
\label{prop:gen-gentle}
Suppose that $\rho^x\in \calD(\calH)$ is a density operator, $\calN^x:\calD(\calH) \to \calD(\calH')$ is a quantum channel  for all $x\in \calX$, and $\mathbf{\Lambda} = \{\Lambda^x\}_{x\in \calX}$ is a POVM. Suppose that $P_X$ is a \ac{PMF} over $\calX$ such that $\sum_{x}P_X(x) \tr{\rho^x \Lambda^x} \geq 1- \epsilon$. It then holds that
\begin{align}
 {\norm{\sum_x P_X(x) \pr{\calN^x(\rho^x) - \sum_{x'} \calN^{x'}( \sqrt{\Lambda^{x'}}\rho^x \sqrt{\Lambda^{x'}})  } }}_1 \leq 2\sqrt{\epsilon} + \epsilon.
\end{align}
\end{proposition}
\begin{IEEEproof}
See Appendix~\ref{sec:gen-gentle}.
\end{IEEEproof}
Let $(f, \mathbf{\Lambda})$ be the $(\mc, \epsilon)^{\mathrm{CC}}$ satisfying $f(\wc) = f_1(\wc) \otimes f_2(\wc)$ for all $\wc\in \intseq{1}{\mc}$, where $f_1:\intseq{1}{\mc} \to \calD\pr{\calH_A^{\proddist n_1}}$, $f_2:\intseq{1}{\mc} \to \calD\pr{\calH_A^{\proddist n_2}}$, and $n_1 + n_2 = n$. Let $\ms^{\mathrm{CC}} \eqdef  \ms^{\mathrm{CC}}(f_2, \zeta)$ and $ K \eqdef K^{\mathrm{CC}}(f_2, \zeta)$. Using the same argument as in the proof of Theorem~\ref{th:main-cc}, there exist an encoder function $g_{\wc}:\intseq{1}{\ms^{\mathrm{CC}}} \times \intseq{1}{K} \to \calD\pr{\calH_{A}^{\proddist n_2}}$ and a POVM $\mathbf{\Gamma}_{\wc, s} = \{\Gamma_{\wc, s}^\ws\}_{\ws \in \intseq{1}{\ms^{\mathrm{CC}}}}$ for each message $\wc\in \intseq{1}{\mc}$ such that
\begin{align}
\label{eq:avg-state}
{\norm{\frac{1}{\ms^{\mathrm{CC}} K} \sum_{\ws, s} \calN_{A\to B}^{\proddist n_2}\pr{ g_w(\ws, s)} -  (\calN_{A\to B} \circ \calM_{A\to A} )^{\proddist n_2}(f_2(w))  }}_1 &\leq \zeta,\\
\frac{1}{\ms^{\mathrm{CC}} K}\sum_{s\ws} \tr{\Gamma_{\wc, s}^\ws \calN_{A\to B}^{\proddist n_2}(g_\wc(\ws, s)) } &\geq 1-\zeta.
\end{align}
Let $\ket{\phi^w}_{RA^{n_1}}$ and $\sigma^w_{RB^{n_1}}$ be defined as in Theorem~\ref{th:main-cc-es}. We define $\ms^w \eqdef E_d(\rho_{RB^{n_1}}^w, \ms^{\mathrm{CC}}, \zeta)$  and fix an $(\ms^w, \ms^{\mathrm{CC}}, \rho_{RB^{n_1}}^w, \zeta)^{\mathrm{ED}}$ protocol $(\calE^w_{R\to C\tA}, \calD^w_{CB^{n_1}\to \tB})$. Let $S_1$ and $S_2$ be two shared secret keys between Alice and Bob uniformly distributed over $\intseq{1}{\ms^{\mathrm{CC}}}$ and $\intseq{1}{K}$, respectively. We define a POVM $\overline{\mathbf{\Lambda}}_{s_2} =\{\overline{\Lambda}^{\wc, \ws}_{s_2}\}_{\wc, \ws} $ with $\overline{\Lambda}^{\wc, \ws}_{s_2} \eqdef \sqrt{\Lambda^\wc}(\one_{B^{n_1}} \otimes \Gamma_{\wc, s_2}^\ws) \sqrt{\Lambda^\wc}$. The stego protocol would operate as follows when $W=\wc$.

 Alice prepares $\ket{\phi^w}_{RA^{n_1}}$ and sends $\phi^w_{A^{n_1}}$ over $n_1$ uses of $\calN_{A\to B}$. Alice then applies $\calE^w_{R\to C\tA}$ to $\phi^w_R$ and sends $g_w(C \oplus S_1, S_2)$ over $n_2$ uses of $\calN_{A\to B}$. Bob performs the POVM $\overline{\mathbf{\Lambda}}_{S_2}$ to decode $W$ and $C$ with the help of $S_1$. Bob finally applies $\calD^w_{CB^{n_1} \to \tB }$ to his first $n_1$ received subsystem to obtain the entangled state.
 
Let  $\rho_{B^n}^\wc$ denote the state received by Bob when $W=\wc$ and  the cover protocol is executed over $n$ uses of $\calN_{A\to B} \circ \calM_{A \to B}$.  Let  $\overline{\rho}_{B^n}^\wc$ denote the state received by Bob when $W=\wc$ and  the stego protocol is executed over $n$ uses of $\calN_{A\to B} $.  Note that both $\rho_{B^n}^w$  and $\overline{\rho}_{B^n}^w$ decompose as
\begin{align}
\rho_{B^n}^w &= (\calN_{A\to B} \circ \calM_{A\to A} )^{\proddist n_1}(f_1(w)) \otimes (\calN_{A\to B} \circ \calM_{A\to A} )^{\proddist n_2}(f_2(w)) \\
\overline{\rho}_{B^n}^w &= \calN_{A\to B}^{\proddist n_1}(\phi^w_{A^{n_1}}) \otimes \calN_{A\to B} ^{\proddist n_2}\pr{\sum_{\ws, s_2} \P{C+S_1 = \ws, S_2 = s_2} g_w(\ws, s_2)}\\
&= \calN_{A\to B}^{\proddist n_1}(\phi^w_{A^{n_1}}) \otimes \pr{  \frac{1}{\ms^{\mathrm{CC} }K}\sum_{\ws s_2}\calN_{A\to B} ^{\proddist n_2}\pr{g_w(\ws, s_2)}}.
\end{align}
We have $(\calN_{A\to B} \circ \calM_{A\to A} )^{\proddist n_1}(f_1(w)) =\calN_{A\to B}^{\proddist n_1}(\phi^w_{A^{n_1}}) $ because $\ket{\phi^w}_{RA^{n_1}}$ is a purification of $\calM_{A\to A} ^{\proddist n_1}(f_1(w))$. We therefore have
\begin{align}
{\norm{\rho_{B^n}^w - \overline{\rho}_{B^n}^w}}_1
&= {\norm{ (\calN_{A\to B} \circ \calM_{A\to A} )^{\proddist n_2}(f_2(w)) -\frac{1}{\ms^{\mathrm{CC} }K}\sum_{\ws s_2}\calN_{A\to B} ^{\proddist n_2}\pr{g_w(\ws, s_2)} }}_1,
\end{align}
which is less than $\zeta$ by \eqref{eq:avg-state}. By the convexity of trace norm, it holds that ${\norm{\rho_{B^n}^c - \rho_{B^n}^s}}_1\leq \zeta$. Following the same reasoning of the proof of Theorem~\ref{th:main-cc}, we conclude that
\begin{align}
\label{eq:decode-prob-th2}
\frac{1}{\mc\ms^{\mathrm{CC}} K  }\frac{1}{\ms^{\mathrm{CC}}, \mc} \sum_{\wc, \ws, , s_2} \tr{\overline{\Lambda}_{s_1}^{\wc, \ws} \calN^{\pn}_{A\to B}(\phi_{A^{n_1}}^\wc \otimes g_\wc(\ws, s_2))} \geq 1- \zeta - 2\sqrt{\epsilon + \zeta}.
\end{align}
In other words, Bob correctly decodes $W$ and $C$ with probability at least $1- \zeta - 2\sqrt{\epsilon + \zeta}$.

 We fix $W=\wc$, $S_1 = s_1$, and $S_2 = s_2$ and denote
\begin{align}
(\calE^\wc_{R\to C\tA}\otimes \id_{A^{n_1}})(\phi_{RA^{n_1}}^\wc) = \sum_c \P{C=c|W=\wc}  \kb{c}_C \otimes \phi_{\tA A^{n_1}}^{\wc, c}.
\end{align} 
 Fixing a value $C=c$ and setting $\ws \eqdef c \oplus s_1$, Alice transmits the subsystem $A^n$ of $\phi_{\tA A^{n_1}}^{\wc, c}\otimes g_\wc(\ws, s_2)$ over $\calN_{A\to B}^{\pn}$, which results in the state $\phi_{\tA B^n}^{\wc, c}$. 

 The shared entangled state would be
\begin{align}
\sum_{\wc'\ws'} \id_{\tA} \otimes \calD^{\wc'}_{CB^{n_1}\to \tB} \otimes \emptyset_{B^{n_2}} \pr{\kb{\ws'-s_1} \otimes  (\one_{\tA} \otimes \sqrt{\overline{\Lambda}_{s_2}^{\wc'\ws'}} )\phi_{\tA B^n}^{\wc, c} (\one_{\tA} \otimes \sqrt{\overline{\Lambda}_{s_2}^{\wc'\ws'}} )}
\end{align}
By \eqref{eq:decode-prob-th2} and Proposition~\ref{prop:gen-gentle}, we obtain that
\begin{multline}
\left\| \sum_{\wc c s_2} \P{W = \wc, C = c, S_2 = s_2}\right.\\
\left. \times \sum_{\wc'\ws'} \id_{\tA} \otimes \calD^{\wc'}_{CB^{n_1}\to \tB} \otimes \emptyset_{B^{n_2}}\pr{\kb{\ws'-s_1} \otimes  (\one_{\tA} \otimes \sqrt{\overline{\Lambda}_{s_2}^{\wc'\ws'}} )\phi_{\tA B^n}^{\wc, c} (\one_{\tA} \otimes \sqrt{\overline{\Lambda}_{s_2}^{\wc'\ws'}} )}  \right.\\
-\left.\sum_{\wc c s_2} \P{W = \wc, C = c, S_2 = s_2}\id_{\tA} \otimes \calD^{\wc}_{CB^{n_1}\to \tB} \otimes \emptyset_{B^{n_2}} \pr{\kb{c} \otimes  \phi_{\tA B^n}^{\wc, c}}\right\|_1 \\\leq \zeta + 2\sqrt{\epsilon + \zeta} + 2\sqrt{\zeta + 2\sqrt{\epsilon + \zeta}}.
\end{multline}
By the definition of an entanglement distillation code, we have 
\begin{align}
{\norm{\sum_c \P{C=c|W =\wc} \id_{\tA} \otimes \calD_{CB^{n_1} \to \tB}(\kb{c}_C \otimes \phi_{\tA B^{n_1}}^{\wc, c} )- \Phi^{(\ms)}} }_1 \leq \zeta.
\end{align}
Using the triangle inequality completes the proof.


\end{IEEEproof}

\begin{IEEEproof}[Proof of Theorem~\ref{th:main-es}]
Let $\ket{\phi}_{R\tA A^n}$ be a purification of $\rho_{\tA A^n}$. Let $V_{A\to BE}$ and $W_{B^n \to \tB H}$ be isometric extensions of $\calN_{A\to B}\circ\calM_{A\to A} =\calM_{A\to A} $ and  $\calD_{B^n \to \tB}$, respectively.  The stego protocol will be as follows. Alice prepares a pure state 
$\ket{\omega}_{R\tA B^n E^n} \eqdef \one_{R\tA} \otimes V_{A\to BE}^\pn \ket{\phi}_{R \tA A^n}$ and sends $\omega_{B^n}$ over $\calN_{A\to B}^\pn$. Bob applies $W_{B^n \to \tB H}$ on $\omega_{B^n}$, which results in the overall state
\begin{align}
\label{eq:psi-def}
\ket{\psi}_{R\tA E^n  \tB H} \eqdef (\one_{R\tA E^n} \otimes W_{B^n\to \tB H}) \circ (\one_{R \tA} \otimes V_{A\to B E}^\pn) \ket{\phi}_{R\tA A^n},
\end{align}
Note that  $F(\Phi^{(M)}, \psi_{\tA \tB}) \geq 1-\epsilon$ from our assumption on the code. We now follow a standard application of Uhlmann's theorem to show that Bob can indeed decode  $\psi_{R E}$. Note that $\ket{\psi}_{R\tA E^n \tB H}$ is a purification of $\psi_{\tA \tB}$. The state $\Phi^{(M)}$ also has a purification over $\calH_{R\tA E^n  \tB H}$. Uhlmann's theorem therefore implies the existence of a purification $\ket{\tau}_{R\tA E^n  \tB H}$  of $\Phi^{(M)}$ over $\calH_{R\tA E^n  \tB H}$ such that $|\braket{\tau}{\psi}_{R\tA E^n  \tB H}|^2 = F(\Phi^{(M)}, \psi_{\tA \tB})$. The vector $\ket{\Phi^{(M)}} \otimes \ket{0}$ is another purification of $ \Phi^{(M)}$ for every unit vector $\ket{0}\in \calH_{R E^n  H}$. By \cite{wilde2013quantum}, there exists a unitary $T_{R E^n H \to R E^nH}$ on $\calH_{R E^n  H}$ such that 
\begin{align*}
\ket{\tau}_{R\tA E^n  \tB H} = (\one_{\tA \tB} \otimes T_{R E^n  H \to R E^n  H}) \ket{\Phi^{(M)}}_{\tA \tB} \otimes \ket{0} = \ket{\Phi^{(M)}}  \otimes (T_{R E^nH \to R E^nF^n  H}\ket{0}) \eqdef  \ket{\Phi^{(M)}}  \otimes \ket{\tau'}_{R E^n  H}.
\end{align*}
We thus have $\braket{\tau}{\psi}_{R\tA E^n \tB H}
=(\bra{ \Phi^{(M)}} \otimes \bra{\tau}_{R E^n  H})\ket{\psi}_{R\tA E^n  \tB H}.$
 We consider a Schmidt decomposition of $\ket{\tau'}_{R E^n  H}$ such as $\ket{\tau'}_{R E^n  H} = \sum_{x\in \calX}\sqrt{P_X(x)} \ket{\alpha_x}_{R E^n}\otimes \ket{\beta_x}_{H}$, where $P_X$ is a \ac{PMF} over $\calX$, and $\ket{\alpha_x}_{R E^n}$ and $\ket{\beta_x}_H$ are orthonormal in $RE^n$ and $H$, respectively. Let $X$ be a random variable distributed according to $P_X$, $f:\calX \to \intseq{1}{\ms}$, and $Q$ and $P_{f(X)}$ be the uniform distribution and the distribution of $f(X)$, respectively. By \cite[Corollary 5.6.1]{renner2008security} and  \cite[Theorem 7]{Tomamichel2009}, when $\log \ms \leq \sup_{a\in]0, 1[} H^a(P_X) - \frac{4}{a}\log \frac{2}{\zeta}$, there exists a function  $f$ such that ${\norm{Q- P_{f(X)}}}_1 \leq \zeta$. Alice measures $\{\Lambda^\ws_A \eqdef \sum_{x:f(x) = \ws} \ket{\alpha_x}\bra{\alpha_x}_{RE^n} \}_{\ws \in\intseq{1}{\ms}}$ on $\psi_{RE^n}$ and Bob measures $\{\Lambda^\ws_B \eqdef \sum_{x:f(x) = \ws} \ket{\beta_x}\bra{\beta_x}_{H} \}_{\ws \in\intseq{1}{\ms}}$ on $\psi_H$. Let $W_A$ and $W_B$ denote the output of the Alice's and Bob's measurement, respectively and $\calA_{RE^n \to W_A}$ and $\calB_{H\to W_B}$ denote the corresponding quantum channels to these measurements.  We have
\begin{align}
&\sqrt{F((\id_{\tA \tB} \otimes\calA_{RE^n \to W_A}\otimes \calB_{H\to W_B})(\psi_{\tA \tB RE^n H}),  \Phi^{(M)} \otimes  \overline{\Phi}^{(\ms)} )}\displaybreak[0]\\
& \geq 1 - {\norm{(\id_{\tA \tB} \otimes\calA_{RE^n \to W_A}\otimes \calB_{H\to W_B})(\psi_{\tA \tB RE^n H})-  \Phi^{(M)} \otimes \overline{\Phi}^{(\ms)}}}_1\displaybreak[0]\\
& \geq 1 - \left\|(\id_{\tA \tB} \otimes\calA_{RE^n \to W_A}\otimes \calB_{H\to W_B})(\psi_{\tA \tB RE^n H})-(\id_{\tA \tB} \otimes\calA_{RE^n \to W_A}\otimes \calB_{H\to W_B})( \Phi^{(M)}\right.\displaybreak[0]\nonumber\\
&\left. \otimes \ket{\tau'}\bra{\tau'}_{RE^nH})\right\|_1
 -{\norm{(\id_{\tA \tB} \otimes\calA_{RE^n \to W_A}\otimes \calB_{H\to W_B})( \Phi^{(M)}\otimes \ket{\tau'}\bra{\tau'}_{RE^nH})-  \Phi^{(M)} \otimes  \overline{\Phi}^{(\ms)}}}_1\displaybreak[0]\\
&\geq  1 - {\norm{\psi_{\tA \tB RE^n H} -  \Phi^{(M)}\otimes \ket{\tau'}\bra{\tau'}_{RE^nH}}}_1\nonumber\displaybreak[0]\\
&\phantom{======} -{\norm{(\id_{\tA \tB} \otimes\calA_{RE^n \to W_A}\otimes \calB_{H\to W_B})( \Phi^{(M)}\otimes \ket{\tau'}\bra{\tau'}_{RE^nH})- \Phi^{(M)} \otimes  \overline{\Phi}^{(\ms)}}}_1 \displaybreak[0]\\
&\geq  1 - \sqrt{1-F(\psi_{\tA \tB RE^n H},  \Phi^{(M)}\otimes \ket{\tau'}\bra{\tau'}_{RE^nH})}\nonumber\\
&\phantom{======} -{\norm{(\id_{\tA \tB} \otimes\calA_{RE^n \to W_A}\otimes \calB_{H\to W_B})( \Phi^{(M)}\otimes \ket{\tau'}\bra{\tau'}_{RE^nH})-  \Phi^{(M)} \otimes  \overline{\Phi}^{(\ms)}}}_1 \\
&\geq  1 - \sqrt{\epsilon}-{\norm{(\id_{\tA \tB} \otimes\calA_{RE^n \to W_A}\otimes \calB_{H\to W_B})( \Phi^{(M)}\otimes \ket{\tau'}\bra{\tau'}_{RE^nH})-  \Phi^{(M)} \otimes  \overline{\Phi}^{(\ms)}}}_1.
\end{align}
We can also write
\begin{align*}
(\id_{\tA \tB} \otimes\calA_{RE^n \to W_A}\otimes \calB_{H\to W_B})(\Phi^{(\mc)}\otimes \ket{\tau'}\bra{\tau'}_{RE^nH}) 
&= \Phi^{(M)}\otimes \pr{\sum_{x}P_X(x) \ket{f(x)f(x)} \bra{f(x)f(x)}_{W_A W_B}}\\
&=   \Phi^{(M)} \otimes \pr{\sum_{\ws}P_{f(X)}(\ws) \ket{\ws \ws} \bra{\ws\ws}_{W_A W_B}}.
\end{align*}
Hence, 
\begin{align*}
{\norm{(\id_{\tA \tB} \otimes\calA_{RE^n \to W_A}\otimes \calB_{H\to W_B})(\Phi^{(M)}\otimes \ket{\tau'}\bra{\tau'}_{RE^nH})- \Phi^{(M)}\otimes \overline{\Phi}^{(\ms)}}}_1 \leq {\norm{P_{f(W) }-Q}}_1 \leq \zeta .
\end{align*}
\end{IEEEproof}

\begin{IEEEproof}[Proof of Theorem~\ref{th:main-qc}]
We start the proof by a technical lemma that helps us simplify the expression of the rate of the cypher message. Let $(\calE_{W\to A}, \calD_{A\to W})$  be an $(M, \epsilon)_{\textnormal{R}}^{\textnormal{QC}}$ code for one use of the channel $\calM_{A\to A}$. Suppose that $\calE_{W\to A}(\rho_W) = V_{W\to A}\rho_W V_{W\to A}^\dagger$ where $V_{W\to A}$ is an isometry, and $\Pi = V_{W\to A}V_{W\to A}^\dagger$ is the projector on to the range of $V_{W\to A}$. Consider a decomposition, $\calM_{A\to A} = \widetilde{\calM}_{A\to A} + \widetilde{\widetilde{\calM}}_{A\to A}$ such that $\calD_{B\to W}\circ \widetilde{\calM}_{A\to A} \circ \calE_{W\to A} = c~\id_W$ for $c\geq 1-\epsilon$. There exists a Kraus representation $\{F_j\}_{j\in \calJ}$ for $\widetilde{\calM}_{A\to A}$ such that $\Pi F_j^\dagger F_{j'}\Pi = \indic{j=j'}d_j \Pi$ for real positive  numbers $\{d_j:j\in \calJ\}$. Define a \ac{PMF} $P_J$ over $\calJ$ as $P_J(j) \eqdef \frac{d_j}{\sum_{j'}d_{j'}}$.
\begin{lemma}
\label{lm:hmin-pj}
For all $\delta>2\sqrt{\epsilon}$, we have $H_{\min}^{\delta}(P_J) \geq H_{\min}^{\delta- 2\sqrt{\epsilon}}\pr{\calM^c_{A\to A}\pr{\frac{1}{M} \Pi}}$, where $\calM^c_{A\to A}$ is the complementary channel of $\calM_{A\to A}$.
\end{lemma}
\begin{IEEEproof}
See Appendix~\ref{sec:hmin-pj}
\end{IEEEproof}
Consider the $(M, \epsilon)^{\textnormal{QC}}_\textnormal{R}$ cover protocol $(\calE_{W\to A^n}, \calD_{B^n \to W})$ for the channel $ \calM_{A\to A}^\pn$. Let $\calE_{W\to A^n}(\rho) = V_{W\to A^n}\rho V_{W\to A^n}^\dagger$ where  $V_{W\to A^n}$ is  an isometry, and $\Pi$ denote the projector onto the range of $V_{W\to A^n}$. By definition, there exists a decomposition $\calM_{A\to A}^\pn = \widetilde{\calM}_{A^n\to A^n} + \widetilde{\widetilde{\calM}}_{A^n\to A^n}$ such that $ \calD_{B^n \to W} \circ \widetilde{\calM}_{A^n\to A^n} \circ \calE_{W\to A^n} = c~\id_W$ with $c\geq 1-\epsilon$. By the same argument as in the proof of \cite[Theorem 10.1]{nielsen2002quantum}, there exists a Kraus representation $\{F_j\}_{j\in \calJ}$ for $\widetilde{\calM}_{A^n\to A^n}$ such that $\Pi F_{j}^\dagger F_{j'} \Pi = \indic{j = j'} d_j \Pi$. By polar decomposition, we therefore have $F_j\Pi \eqdef U_j \sqrt{\Pi F_j^\dagger F_j\Pi} = \sqrt{d_{j}} U_j \Pi$ for some unitary $U_j$ on $\calH_A^\pn$.

 Let $J$ be distributed according to $P_J(j)\eqdef \frac{d_j}{\sum_{j'}d_{j'}}$, and $Q$ denote the uniform distribution over $\intseq{1}{\ms}$. By \cite[Corollary 5.6.1]{renner2008security}, there exists a function  $g:{\calJ}\to \intseq{1}{\ms}$ such that ${\norm{P_{g({J})} - Q}}_1 \leq \zeta$, provided that 
\begin{align} 
 \log \ms
 & = H_{\min}^{\zeta/2}(P_J)- 2\log \frac{2}{\zeta}\\
 &\stackrel{(a)}{\geq}  H_{\min}^{\zeta/2 - 2\sqrt{\epsilon}}\pr{{\calM^c_{A\to A}}^\pn\pr{\frac{1}{M}\Pi}}- 2\log \frac{2}{\zeta}\\
 &\stackrel{(b)}{\geq} \sup_{a\in]0, 1[} H^a\pr{{\calM^c_{A\to A}}^\pn\pr{\frac{1}{M}\Pi}}- 2\log \frac{2}{\epsilon} - \frac{1}{a} \log\frac{2}{(\zeta/2-2\sqrt{\epsilon})^2}\\
 &\geq \sup_{a\in]0, 1[} H^a\pr{{\calM^c_{A\to A}}^\pn\pr{\frac{1}{M}\Pi}} - \frac{4}{a} \log\frac{2}{\zeta/2-2\sqrt{\epsilon}},
  \end{align}
 where $(a)$ follows from Lemma~\ref{lm:hmin-pj}, and $(b)$ follows from  \cite[Theorem 7]{Tomamichel2009}.

 Let $\mu_\ws \eqdef \sum_{j\in \widetilde{\calJ}} \indic{g(j) = \ws} $. We then define $\overline{\calE}^\ws_{W\to A^n}(\rho) \eqdef  \frac{1}{\mu_\ws} \sum_{j:g(j)=\ws } U_j \calE_{W\to A^n}(\rho) U_j^\dagger$ (for $\mu_\ws = 0$ take $\overline{\calE}^\ws_{W\to A^n} = \calE_{W\to A^n}$).  We define the decoder for Bob as 
\begin{align}
\overline{\calD}_{B^n\to W\overline{W}}(\rho_{B^n}) \eqdef  (\calD_{B^n\to W} \otimes \id_{\overline{W}})\pr{\sum_{j} (PU_j^\dagger\otimes \ket{g(j)}) \rho_{B^n} (U_jP \otimes \bra{g(j)}) + E\rho_{B^n} E^\dagger},
\end{align}
where the term $E\rho_{B^n}^\dagger E^\dagger$ is added to ensure that  $\overline{\calD}_{B^n\to W\overline{W}}$ is trace-preserving. By the argument in the proof of \cite[Theorem 10.1]{nielsen2002quantum}, $\overline{\calD}_{B^n\to W\overline{W}}$ is a valid quantum channel. The partial channels are
\begin{align}
\overline{\calD}_{B^n\to W}(\rho_{B^n}) \eqdef  \calD_{B^n\to W}\pr{\sum_{j} (PU_j^\dagger) \rho_{B^n} (U_jP ) + E'\rho_{B^n} {E'}^\dagger}\displaybreak[0],\\
\overline{\calD}_{B^n\to \overline{W}}(\rho_{B^n}) \eqdef \sum_{j} \tr{PU_j^\dagger \rho_{B^n}U_jP} \ket{g(j)}\bra{g(j)} + E''\rho_{B^n} {E''}^\dagger,\displaybreak[0]
\end{align}

 Furthermore, for any $\rho_W\in \calD(\calH_W)$, we have 
\begin{align}
&\frac{1}{\ms}\sum_{\ws = 1}^\ms \tr{\ket{\ws}\bra{\ws}\calD_{B^n\to \overline{W}}(\calE^\ws_{W\to A^n}(\rho_W))}\displaybreak[0]\\
&= \frac{1}{\ms}\sum_{\ws = 1}^\ms \tr{\ket{\ws}\bra{\ws}\pr{\sum_{j} \tr{PU_j^\dagger \calE_{W\to A^n}^\ws(\rho_W)U_jP} \ket{g(j)}\bra{g(j)} + E''\calE_{W\to A^n}^\ws(\rho_W){E''}^\dagger}}\displaybreak[0]\\
&\geq \frac{1}{\ms}\sum_{\ws = 1}^\ms \tr{\ket{\ws}\bra{\ws}\pr{\sum_{j} \tr{PU_j^\dagger \calE_{W\to A^n}^\ws(\rho_W)U_jP} \ket{g(j)}\bra{g(j)} }}\displaybreak[0]\\
&= \frac{1}{\ms}\sum_{\ws = 1}^\ms \sum_{j:g(j) = \ws} \tr{PU_j^\dagger \calE_{W\to A^n}^\ws(\rho_W)U_jP}\displaybreak[0]\\
&\geq \frac{1}{\ms}\sum_{\ws = 1}^\ms \sum_{j:g(j) = \ws} \indic{\mu_\ws \neq 0} \tr{PU_j^\dagger\pr{\frac{1}{\mu_\ws} \sum_{j':g(j')=\ws } U_{j'} \calE_{W\to A^n}(\rho) U_{j'}^\dagger}U_jP}\displaybreak[0]\\
&\geq \frac{1}{\ms}\sum_{\ws = 1}^\ms \frac{1}{\mu_\ws} \sum_{j:g(j) = \ws} \indic{\mu_\ws \neq 0} \tr{PU_j^\dagger U_{j} \calE_{W\to A^n}(\rho) U_{j}^\dagger U_jP}\displaybreak[0]\\
&= \tr{P\calE_{W\to A^n}(\rho_W) P} \frac{1}{\ms} \sum_{\ws} \indic{\mu_\ws \neq 0}\displaybreak[0]=  \frac{1}{\ms} \sum_{\ws} \indic{\mu_\ws \neq 0} \displaybreak[0]\geq 1- {\norm{P_{g(J)} - Q}}_1 \geq 1-\zeta.\nonumber
\end{align}
For a $\ws\in \intseq{1}{\ms}$ we have 
\begin{align*}
&\overline{\calD}_{B^n \to W} \circ \widetilde{\calM}_{A^n\to A^n} \circ \calE_{W\to A^n}^\ws (\rho_W)\displaybreak[0]\\
&= \calD_{B^n\to W}\pr{\sum_{j} (PU_j^\dagger)\widetilde{\calM}_{A^n\to A^n}\circ \calE_{W\to A^n}^\ws (\rho_W)  (U_jP ) + E'\widetilde{\calM}_{A^n\to A^n}\circ \calE_{W\to A^n}^\ws (\rho_W) {E'}^\dagger}\displaybreak[0] \\
&= \calD_{B^n\to W}\pr{\sum_{j} (PU_j^\dagger)\widetilde{\calM}_{A^n\to A^n}\circ \calE_{W\to A^n}^\ws (\rho_W)  (U_jP ) }.
\end{align*}
We can write 
\begin{align*}
\sum_{j} (PU_j^\dagger)\widetilde{\calM}_{A^n\to A^n}\circ \calE_{W\to A^n}^\ws (\rho_W)  (U_jP ) 
&= \sum_{j} (PU_j^\dagger)\widetilde{\calM}_{A^n\to A^n}\pr{ \frac{1}{\mu_\ws} \sum_{j':g(j')=\ws } U_{j'} \calE_{W\to A^n}(\rho_W) U_{j'}^\dagger}(U_jP ) \\
&=\widetilde{\calM}_{A^n\to A^n}\calE_{W\to A^n}(\rho_W).
\end{align*}
Hence, it holds that $\overline{\calD}_{B^n \to W} \circ \widetilde{\calM}_{A^n\to A^n} \circ \calE_{W\to A^n}^\ws  = \calD_{B^n\to W} \circ \widetilde{\calM}_{A^n\to A^n} \circ  \overline{\calE}_{W\to A^n} = c \id_W$ for $c\geq 1-\epsilon$. Finally for all $\rho_W$, we have
\begin{align*}
{\norm{\rho_{B^n}^c - \rho_{B^n}^s}}_1
&={\norm{\calM^\pn_{A\to A}\circ\calE_{W\to A^n} (\rho_W)  - \frac{1}{\ms} \sum_{\ws}\overline{\calE}_{W\to A^n}^\ws (\rho_W) }}_1\\
& = {\norm{\widetilde{\calM}_{A^n\to A^n}\circ\calE_{W\to A^n} (\rho_W)  - \frac{1}{\ms} \sum_{\ws}\overline{\calE}_{W\to A^n}^\ws (\rho_W) }}_1+ {\norm{\widetilde{\widetilde{\calM}}_{A^n\to A^n}\circ \calE_{W\to A^n}(\rho_W)}}_1.
\end{align*}
For the first term, we have
\begin{align}
&{\norm{\widetilde{\calM}_{A^n\to A^n}\circ\calE_{W\to A^n} (\rho_W)  - \frac{1}{\ms} \sum_{\ws}\overline{\calE}_{W\to A^n}^\ws (\rho_W) }}_1\displaybreak[0] \\
&\phantom{========}= {\norm{\sum_{j\in \calJ} F_j\calE_{W\to A^n} (\rho_W)F_j^\dagger  - \frac{1}{\ms} \sum_{\ws}\frac{1}{\mu_\ws} \sum_{j:g(j)=\ws} U_j \calE_{W\to A^n} (\rho_W)U_j^\dagger  }}_1\displaybreak[0]\\
&\phantom{========}={\norm{\sum_{j\in \calJ} F_j\calE_{W\to A^n} (\rho_W)F_j^\dagger  - \sum_{j\in \calJ} P_{g(J)}(j) U_j \calE_{W\to A^n} (\rho_W)U_j^\dagger  }}_1\displaybreak[0]\\
&\phantom{========}\stackrel{(a)}{=}{\norm{\sum_{j\in \calJ} d_j U_j\calE_{W\to A^n} (\rho_W)U_j^\dagger  - \sum_{j\in \calJ} P_{g(J)}(j) U_j \calE_{W\to A^n} (\rho_W)U_j^\dagger  }}_1\displaybreak[0]\\
&\phantom{========}\leq \sum_{j\in \calJ}|d_j - P_{g(J)}(j)|\displaybreak[0]\\
&\phantom{========}\leq \sum_{j\in \calJ}\left|d_j - \frac{d_j}{\sum_{j'\in \calJ} d_{j'}}\right| +\sum_{j\in \calJ}\left| \frac{d_j}{\sum_{j'\in \calJ} d_{j'}}- P_{g(J)}(j)\right| \leq \epsilon + \zeta\displaybreak[0],
\end{align}
where $(a)$ follows since by the definition of $\calE_{W\to A^n}$, $\Pi \calE_{W\to A^n}(\rho_W) \Pi = \calE_{W\to A^n}(\rho_W)$, and $F_j \Pi = \sqrt{d_j} U_j \Pi$.
Furthermore,
\begin{align}
 {\norm{\widetilde{\widetilde{\calM}}_{A^n\to A^n}\circ \calE_{W\to A^n}(\rho_W)}}_1 
 & =  \tr{\widetilde{\widetilde{\calM}}_{A^n\to A^n}\circ \calE_{W\to A^n}(\rho_W)}\displaybreak[0]\\
  & =  \tr{\calD_{B^n\to W} \circ \widetilde{\widetilde{\calM}}_{A^n\to A^n}\circ \calE_{W\to A^n}(\rho_W)}\displaybreak[0]\\
    & =  \tr{\calD_{B^n\to W} \circ \pr{\calM_{A\to A}^\pn-{\widetilde{\calM}}_{A^n\to A^n}}\circ \calE_{W\to A^n}(\rho_W)}\displaybreak[0]\\
    &= 1-c\leq \epsilon.
\end{align}
\end{IEEEproof}

\appendices
\section{Proof of Proposition~\ref{prop:gen-gentle}}
\label{sec:gen-gentle}
By the triangle inequality, we have
\begin{align}
&{\norm{\sum_x P_X(x) \pr{\calN^x(\rho^x) - \sum_{x'} \calN^{x'}(\sqrt{\Lambda^{x'}}\rho^x \sqrt{\Lambda^{x'}})  } }}_1\displaybreak[0]\\
&\leq {\norm{\sum_x P_X(x) \pr{\calN^x(\rho^x) - \calN^{x}(\sqrt{\Lambda^{x}}\rho^x \sqrt{\Lambda^{x}}) } }}_1 + {\norm{\sum_{x\neq x'} P_X(x) \calN^{x'}(\sqrt{\Lambda^{x'}}\rho^x \sqrt{\Lambda^{x'}})  }}_1.\displaybreak[0]
\end{align}
Since $\calN^{x'}(\sqrt{\Lambda^{x'}}\rho^x \sqrt{\Lambda^{x'}})$ is positive semi-definite, the second term would simplify as
\begin{align}
{\norm{\sum_{x\neq x'} P_X(x) \calN^{x'}(\sqrt{\Lambda^{x'}}\rho^x \sqrt{\Lambda^{x'}})  }}_1 \displaybreak[0]
&= \sum_{x\neq x'} P_X(x) \tr{\calN^{x'}(\sqrt{\Lambda^{x'}}\rho^x \sqrt{\Lambda^{x'}})}\\\displaybreak[0]
&= \sum_{x\neq x'} P_X(x) \tr{\sqrt{\Lambda^{x'}}\rho^x \sqrt{\Lambda^{x'}}}\\\displaybreak[0]
&= \sum_{x\neq x'} P_X(x) \tr{{\Lambda^{x'}}\rho^x}\\\displaybreak[0]
&= 1- \sum_x P_X(x) \tr{\Lambda^x \rho^x} \leq \epsilon.
\end{align}
Furthermore, by the gentle measurement lemma, we have
\begin{align}
{\norm{\sum_x P_X(x) \pr{\calN^x(\rho^x) - \calN^x(\sqrt{\Lambda^{x}}\rho^x \sqrt{\Lambda^{x}})  } }}_1
&\leq \sum_x P_X(x)  {\norm{{\calN^x(\rho^x) - \calN^x(\sqrt{\Lambda^{x}}\rho^x \sqrt{\Lambda^{x}})  } }}_1\displaybreak[0]\\
&\stackrel{(a)}{\leq}\sum_x P_X(x)  {\norm{{\rho^x - \sqrt{\Lambda^{x}}\rho^x \sqrt{\Lambda^{x}}  } }}_1\displaybreak[0]\\
&\leq2 \sum_x P_X(x)\sqrt{1- \tr{\Lambda^x \rho^x}}\displaybreak[0] \\
&\stackrel{(b)}{\leq} 2 \sqrt{1- \sum_x P_X(x)\tr{\Lambda^x \rho^x}\displaybreak[0]} \leq 2\sqrt{\epsilon},
\end{align}
where $(a)$ follows from the date processing inequality (which holds for non-normalized states), and $(b)$ follows from the concavity of the mapping $x\mapsto \sqrt{1-x}$.
\section{Proof of Lemma~\ref{lm:hmin-pj}}
\label{sec:hmin-pj}
 We extend $\{F_j\}_{j\in\calJ}$ to a Kraus representation $\{F_{j}\}_{j\in \calK}$ for the channel $\calM_{A\to A}$ with $\calJ \subset \calK$. By \cite{wilde2013quantum}, $U_{A\to AE} \eqdef \sum_{j\in \calK} F_j \otimes \ket{j}_E$ is an isometric extension of $\calM_{A\to A}$ where $\{\ket{j}_E\}_{j\in \calK}$ is an orthonormal basis for the environment space $\calH_E$. Let $\rho_A \eqdef \calE_{W\to A}\pr{\frac{1}{M}\one_W} = \frac{1}{M} \Pi$, and 
\begin{align}
\rho_E 
&\eqdef \textnormal{tr}_A(U_{A\to AE} \rho_A U_{A\to AE} ^\dagger)\displaybreak[0]\\
&= \textnormal{tr}_A\pr{\pr{\sum_{j\in \calK} F_j \otimes \ket{j}_E} \pr{\frac{1}{M} \Pi}\pr{\sum_{j\in \calK} F_j \otimes \ket{j}_E}^\dagger}\displaybreak[0]\\
&= \frac{1}{M} \sum_{j, j'\in \calK } \tr{F_j \Pi F_{j'}^\dagger} \ket{j}\bra{j'}_E.\displaybreak[0]
\end{align} 
For the projector $\Gamma \eqdef \sum_{j\in \calJ} \ket{j}\bra{j}_E$, we have
\begin{align}
\tr{\Gamma \rho_E} \displaybreak[0]
&= \tr{\pr{\sum_{j\in \calJ} \ket{j}\bra{j}_E}\pr{\frac{1}{M} \sum_{j, j'\in \calK } \tr{F_j \Pi F_{j'}^\dagger} \ket{j}\bra{j'}_E}}\\\displaybreak[0]
&=\frac{1}{M} \sum_{j \in \calJ} \tr{F_j \Pi F_{j}^\dagger}\\\displaybreak[0]
&\stackrel{(a)} =  \tr{\widetilde{\calM}_{A\to A}\pr{\frac{1}{M} \Pi}}\\\displaybreak[0]
&= \tr{\widetilde{\calM}_{A\to A}\pr{\calE_{W\to A}\pr{\frac{1}{M}\one_W}}}\\
&\stackrel{(b)}{=} \tr{\calD_{A\to W}\pr{\widetilde{\calM}_{A\to A}\pr{\calE_{W\to A}\pr{\frac{1}{M}\one_W}}}}\\
&= \tr{{c \id_W\pr{\frac{1}{M}\one_W}}}\\
&=c \geq 1-\epsilon,
\end{align}
where $(a)$ follows since $\{F_j\}_{j\in\calJ}$ is a Kraus representation of $\widetilde{\calM}_{A\to A}$, $(b)$ follows since $\calD_{A\to W}$ is trace-preserving. By the gentle measurement lemma \cite{Winter1999}, we obtain that ${\norm{\frac{\Gamma \rho_E \Gamma}{\tr{\Gamma \rho_E}} - \rho_E}}_1 \leq 2\sqrt{\epsilon}$.
We can write 
\begin{align}
\frac{\Gamma \rho_E \Gamma}{\tr{\Gamma \rho_E}}  
&= \frac{1}{\tr{\Gamma \rho_E}} \pr{\sum_{j\in \calJ} \ket{j}\bra{j}_E}\pr{\frac{1}{M} \sum_{j, j'\in \calK } \tr{F_j \Pi F_{j'}^\dagger} \ket{j}\bra{j'}_E}\pr{\sum_{j\in \calJ} \ket{j}\bra{j}_E}\\
&= \frac{1}{\tr{\Gamma \rho_E}M}  \sum_{j, j'\in \calJ} \tr{F_j \Pi F_{j'}\dagger} \ket{j}\bra{j'}_E= \frac{1}{\tr{\Gamma \rho_E}M}  \sum_{j}d_j \ket{j}\bra{j}_E= \sum_{j\in \calJ} P_J(j) \ket{j}\bra{j}_E.\nonumber
\end{align}
Therefore, we have $H_{\min}^{\delta}(P_J) \geq H_{\min}^{\delta -2\sqrt{\epsilon}}(\rho_E)$
\bibliographystyle{IEEEtran}
\bibliography{quantumstegano}
\end{document}